\documentclass[12pt]{article}
\usepackage{graphicx}
\usepackage{subfigure}
\usepackage[nosort]{cite}
\usepackage{color,url,hyperref}
\usepackage{caption}
\newcommand{\newc}{\newcommand}

\def\Ord{\lower .7ex\hbox{$\;\stackrel{\textstyle <}{\sim}\;$}}
\def\OOrd{\lower .7ex\hbox{$\;\stackrel{\textstyle >}{\sim}\;$}}

\newc{\order}{{\cal O}}

\newc{\be}{\begin{equation}}
\newc{\ee}{\end{equation}}
\newc{\br}{\begin{eqnarray}}
\newc{\er}{\end{eqnarray}}
\newc{\ba}{\begin{array}}
\newc{\ea}{\end{array}}
\newc{\bi}{\begin{itemize}}
\newc{\ei}{\end{itemize}}
\newc{\bn}{\begin{enumerate}}
\newc{\en}{\end{enumerate}}
\newc{\bc}{\begin{center}}
\newc{\ec}{\end{center}}
\newc{\ul}{\underline}
\newc{\ra}{\rightarrow}
\newc{\lra}{\longrightarrow}
\newc{\wt}{\widetilde}
\newc{\til}{\tilde}

\newc{\wh}{\widehat}
\newc{\ti}{\times}
\newc{\Dir}{\kern -6.4pt\Big{/}}
\newc{\Dirin}{\kern -10.4pt\Big{/}\kern 4.4pt}
\newc{\DDir}{\kern -10.6pt\Big{/}}
\newc{\DGir}{\kern -6.0pt\Big{/}}
\newc{\sig}{\sigma}
\newc{\sigmalstop}{\sig_{\lstoppair}}
\newc{\Sig}{\Sigma}  
\newc{\del}{\delta}
\newc{\Del}{\Delta}
\newc{\lam}{\lambda}
\newc{\Lam}{\Lambda}
\newc{\gam}{\gamma}
\newc{\Gam}{\Gamma}
\newc{\eps}{\epsilon}
\newc{\Eps}{\Epsilon}
\newc{\kap}{\kappa}
\newc{\Kap}{\Kappa}
\newc{\modulus}[1]{\left| #1 \right|}
\newc{\eq}[1]{(\ref{eq:#1})}
\newc{\eqs}[2]{(\ref{eq:#1},\ref{eq:#2})}
\newc{\etal}{{\it et al.}\ }
\newc{\ibid}{{\it ibid}.}
\newc{\ibidem}{{\it ibidem}.}
\newc{\eg}{{\it e.g.}\ }
\newc{\ie}{{\it i.e.}\ }

\newc{\nonum}{\nonumber}
\newc{\lab}[1]{\label{eq:#1}}
\newc{\dpr}[2]{({#1}\cdot{#2})}
\newc{\lt}{\stackrel{<}}
\newc{\gt}{\stackrel{>}}
\newc{\lsimeq}{\stackrel{<}{\sim}}
\newc{\gsimeq}{\stackrel{>}{\sim}}
\def\lsim{\buildrel{\scriptscriptstyle <}\over{\scriptscriptstyle\sim}}
\def\gsim{\buildrel{\scriptscriptstyle >}\over{\scriptscriptstyle\sim}}
\def\lapp{\mathrel{\rlap{\raise.5ex\hbox{$<$}}
                    {\lower.5ex\hbox{$\sim$}}}}
\def\gapp{\mathrel{\rlap{\raise.5ex\hbox{$>$}}
                    {\lower.5ex\hbox{$\sim$}}}}
\newc{\half}{\frac{1}{2}}

\newc{\bQ}{\ol{Q}}
\newc{\dota}{\dot{\alpha }}
\newc{\dotb}{\dot{\beta }}
\newc{\dotd}{\dot{\delta }}
\newc{\nindnt}{\noindent}

\newc{\matth}{\mathsurround=0pt}
\def\ML{\ifmmode{{\mathaccent"7E M}_L}
             \else{${\mathaccent"7E M}_L$}\fi}
\def\MR{\ifmmode{{\mathaccent"7E M}_R}
             \else{${\mathaccent"7E M}_R$}\fi}

\newc{\mr}{\mathrm}

\newc{\siminf}{\mbox{$_{\sim}$ {\small {\hspace{-1.em}{$<$}}}    }}
\newc{\simsup}{\mbox{$_{\sim}$ {\small {\hspace{-1.em}{$>$}}}    }}

\newc {\Zboson}{{\mathrm Z}^{0}}
\newc{\thetaw}{\theta_W}
\newc{\mbot}{{m_b}}
\newc{\mtop}{{m_t}}
\newc{\sm}{${\cal {SM}}$}
\newc{\as}{\alpha_s}
\newc{\aem}{\alpha_{em}}

\newc{\ppbar}{\mbox{$p\ol{p}$}}
\newc{\bbbar}{\mbox{$b\ol{b}$}}
\newc{\ccbar}{\mbox{$c\ol{c}$}}
\newc{\ttbar}{\mbox{$t\ol{t}$}}
\newc{\eebar}{\mbox{$e\ol{e}$}}
\newc{\zzero}{\mbox{$Z^0$}}

\newc{\wplus}{\mbox{$W^+$}}
\newc{\wminus}{\mbox{$W^-$}}
\newc{\ellp}{\ell^+}
\newc{\ellm}{\ell^-}
\newc{\elp}{\mbox{$e^+$}}
\newc{\elm}{\mbox{$e^-$}}
\newc{\elpm}{\mbox{$e^{\pm}$}}
\newc{\qbar}     {\mbox{$\ol{q}$}}

\newc{\Ebar}{{\bar E}}
\newc{\Dbar}{{\bar D}}
\newc{\Ubar}{{\bar U}}
\newc{\susy}{{{SUSY}}}
\newc{\msusy}{{{M_{SUSY}}}}

\def\photino{\ifmmode{\mathaccent"7E \gam}\else{$\mathaccent"7E \gam$}\fi}
\def\taugluino{\ifmmode{\tau_{\mathaccent"7E g}}
             \else{$\tau_{\mathaccent"7E g}$}\fi}
\def\mphotino{\ifmmode{m_{\mathaccent"7E \gam}}
             \else{$m_{\mathaccent"7E \gam}$}\fi}
\newc{\gl}   {\mbox{$\wt{g}$}}
\newc{\mgl}  {\mbox{$m_{\gl}$}}

\def \chonep {{\wt\chi_1^+}}

\def \ch2p {{\wt\chi_2^+}}
\def \chonem {{\wt\chi_1^-}}
\def \ch2m {{\wt\chi_2^-}}

\def \chonepm{{\wt\chi_1}^{\pm}}

\def \mchonepm{m_{\chonepm}}

\def \chtwom{{\wt\chi_2}^{-}}

\newc{\dmchi}{\Delta m_{\wt\chi}}

\def \lspone{\wt\chi_1^0}
\def \mlspone{m_{\lspone}}
\def \lsptwo{\wt\chi_2^0}

\newc{\sele}{\wt{\mathrm e}}
\newc{\sell}{\wt{\ell}}

\newc{\snue}     {\mbox{$ \wt{\nu_e}$}}
\newc{\smu}{\wt{\mu}}
\newc{\stau}{\wt{\tau}}
\newc {\nuL} {\wt{\nu}_L}
\newc {\nuR} {\wt{\nu}_R}
\newc {\snub} {\bar{\wt{\nu}}}
\newc {\eL} {\wt{e}_L}
\newc {\eR} {\wt{e}_R}

\def \stau{\wt\tau}

\def \sq{\wt{q}}

\newc{\msqot}  {\mbox{$m_(\sq_{1,2} )$}}
\newc{\sqbar}    {\mbox{$\bar{\wt{q}}$}}
\newc{\ssb}      {\mbox{$\squark\ol{\squark}$}}
\newc {\qL} {\wt{q}_L}
\newc {\qR} {\wt{q}_R}
\newc {\uL} {\wt{u}_L}
\newc {\uR} {\wt{u}_R}
\def \ul{\wt{u}_L}

\newc {\dL} {\wt{d}_L}
\newc {\dR} {\wt{d}_R}
\newc {\cL} {\wt{c}_L}
\newc {\cR} {\wt{c}_R}
\newc {\sL} {\wt{s}_L}
\newc {\sR} {\wt{s}_R}
\newc {\tL} {\wt{t}_L}
\newc {\tR} {\wt{t}_R}
\newc {\stb} {\ol{\wt{t}}_1}
\newc {\sbot} {\wt{b}_1}
\newc {\msbot} {m_{\sbot}}
\newc {\sbotb} {\ol{\wt{b}}_1}
\newc {\bL} {\wt{b}_L}
\newc {\bR} {\wt{b}_R}

\newc{\csquark}  {\mbox{$\wt{c}$}}
\newc{\csquarkl} {\mbox{$\wt{c}_L$}}
\newc{\mcsl}     {\mbox{$m(\csquarkl)$}}

\newc {\stopl}         {\wt{\mathrm{t}}_{\mathrm L}}
\newc {\stopr}         {\wt{\mathrm{t}}_{\mathrm R}}
\newc {\stoppair}      {\wt{\mathrm{t}}_{1}
\bar{\wt{\mathrm{t}}}_{1}}
\def \lstop{\wt{t}_{1}}

\def \lstoppair{\lstop\lstop^*}

\newc{\tsquark}  {\mbox{$\wt{t}$}}
\newc{\ttb}      {\mbox{$\tsquark\ol{\tsquark}$}}
\newc{\ttbone}   {\mbox{$\tsquark_1\ol{\tsquark}_1$}}

\newc{\mix}{\theta_{\wt t}}
\newc{\cost}{\cos{\theta_{\wt t}}}
\newc{\sint}{\sin{\theta_{\wt t}}}
\newc{\costloop}{\cos{\theta_{\wt t_{loop}}}}

\newc{\mixsbot}{\theta_{\wt b}}

\newc{\tb}{\tan\beta}
\newc{\cb}{\cot\beta}
\newc{\vev}[1]{{\left\langle #1\right\rangle}}

\newc{\mhalf}{m_{1/2}}
\newc{\mzero} {\mbox{$m_0$}}
\newc{\azero} {\mbox{$A_0$}}

\newc{\lb}{\lam}
\newc{\lbp}{\lam^{\prime}}
\newc{\lbpp}{\lam^{\prime\prime}}
\newc{\rpv}{{\not \!\! R_p}}
\newc{\rpvm}{{\not  R_p}}
\newc{\rp}{R_{p}}
\newc{\rpmssm}{{RPC MSSM}}
\newc{\rpvmssm}{{RPV MSSM}}

\newc{\sbyb}{S/$\sqrt B$}
\newc{\pelp}{\mbox{$e^+$}}
\newc{\pelm}{\mbox{$e^-$}}
\newc{\pelpm}{\mbox{$e^{\pm}$}}
\newc{\epem}{\mbox{$e^+e^-$}}
\newc{\lplm}{\mbox{$\ell^+\ell^-$}}

\def\Ecm{\ifmmode{E_{\mathrm{cm}}}\else{$E_{\mathrm{cm}}$}\fi}
\newc{\rts}{\sqrt{s}}
\newc{\rtshat}{\sqrt{\hat s}}
\newc{\gev}{\,GeV}
\newc{\mev}{~{\rm MeV}}
\newc{\tev}  {\mbox{$\;{\rm TeV}$}}
\newc{\gevc} {\mbox{$\;{\rm GeV}/c$}}
\newc{\gevcc}{\mbox{$\;{\rm GeV}/c^2$}}
\newc{\intlum}{\mbox{${ \int {\cal L} \; dt}$}}
\newc{\call}{{\cal L}}
\def \met  {\mbox{${E\!\!\!\!/_T}$}}

\newc{\ptmiss}{/ \hskip-7pt p_T}

\newc{\PT}{\mbox{$p_T$}}
\newc{\ET}{\mbox{$E_T$}}
\newc{\dedx}{\mbox{${\rm d}E/{\rm d}x$}}
\newc{\ifb}{\mbox{${\rm fb}^{-1}$}}
\newc{\ipb}{\mbox{${\rm pb}^{-1}$}}
\newc{\pb}{~{\rm pb}}
\newc{\fb}{~{\rm fb}}
\newc{\ycut}{y_{\mathrm{cut}}}
\newc{\chis}{\mbox{$\chi^{2}$}}

\def \jet(s){\emph{jet(s) }}

\newc{\mpl}{M_{\rm Pl}}
\newc{\mgut}{M_{GUT}}
\newc{\mw}{M_{W}}
\newc{\mweak}{M_{weak}}
\newc{\mz}{M_{Z}}

\newc{\OPALColl}   {OPAL Collaboration}
\newc{\ALEPHColl}  {ALEPH Collaboration}
\newc{\DELPHIColl} {DELPHI Collaboration}
\newc{\XLColl}     {L3 Collaboration}
\newc{\JADEColl}   {JADE Collaboration}
\newc{\CDFColl}    {CDF Collaboration}
\newc{\DXColl}     {D0 Collaboration}
\newc{\HXColl}     {H1 Collaboration}
\newc{\ZEUSColl}   {ZEUS Collaboration}
\newc{\LEPColl}    {LEP Collaboration}
\newc{\ATLASColl}  {ATLAS Collaboration}
\newc{\CMSColl}    {CMS Collaboration}
\newc{\UAColl}    {UA Collaboration}
\newc{\KAMLANDColl}{KamLAND Collaboration}
\newc{\IMBColl}    {IMB Collaboration}
\newc{\KAMIOColl}  {Kamiokande Collaboration}
\newc{\SKAMIOColl} {Super-Kamiokande Collaboration}
\newc{\SUDANTColl} {Soudan-2 Collaboration}
\newc{\MACROColl}  {MACRO Collaboration}
\newc{\GALLEXColl} {GALLEX Collaboration}
\newc{\GNOColl}    {GNO Collaboration}
\newc{\SAGEColl}  {SAGE Collaboration}
\newc{\SNOColl}  {SNO Collaboration}
\newc{\CHOOZColl}  {CHOOZ Collaboration}
\newc{\PDGColl}  {Particle Data Group Collaboration}

\def\issue(#1,#2,#3){{\bf #1}, #2 (#3)}
\def\iss(#1,#2,#3){{\bf #1} (#3) #2}
\def\ASTR(#1,#2,#3){Astropart.\ Phys. \issue(#1,#2,#3)}
\def\AJ(#1,#2,#3){Astrophysical.\ Jour. \issue(#1,#2,#3)}
\def\AJS(#1,#2,#3){Astrophys.\ J.\ Suppl. \issue(#1,#2,#3)}
\def\APP(#1,#2,#3){Acta.\ Phys.\ Pol. \issue(#1,#2,#3)}
\def\JCAP(#1,#2,#3){Journal\ XX. \issue(#1,#2,#3)} 
\def\SC(#1,#2,#3){Science \issue(#1,#2,#3)}
\def\PRD(#1,#2,#3){Phys.\ Rev.\ D \issue(#1,#2,#3)}
\def\PR(#1,#2,#3){Phys.\ Rev.\ \issue(#1,#2,#3)} 
\def\PRC(#1,#2,#3){Phys.\ Rev.\ C \issue(#1,#2,#3)}
\def\NPB(#1,#2,#3){Nucl.\ Phys.\ B \issue(#1,#2,#3)}
\def\NPPS(#1,#2,#3){Nucl.\ Phys.\ Proc. \ Suppl \issue(#1,#2,#3)}
\def\NJP(#1,#2,#3){New.\ J.\ Phys. \issue(#1,#2,#3)}
\def\JP(#1,#2,#3){J.\ Phys.\issue(#1,#2,#3)}
\def\PL(#1,#2,#3){Phys.\ Lett. \issue(#1,#2,#3)}
\def\ZP(#1,#2,#3){Z.\ Phys. \issue(#1,#2,#3)}
\def\ZPC(#1,#2,#3){Z.\ Phys.\ C  \issue(#1,#2,#3)}
\def\PREP(#1,#2,#3){Phys.\ Rep. \issue(#1,#2,#3)}
\def\PRL(#1,#2,#3){Phys.\ Rev.\ Lett. \issue(#1,#2,#3)}
\def\MPL(#1,#2,#3){Mod.\ Phys.\ Lett. \issue(#1,#2,#3)}
\def\RMP(#1,#2,#3){Rev.\ Mod.\ Phys. \issue(#1,#2,#3)}
\def\SJNP(#1,#2,#3){Sov.\ J.\ Nucl.\ Phys. \issue(#1,#2,#3)}
\def\CPC(#1,#2,#3){Comp.\ Phys.\ Comm. \issue(#1,#2,#3)}
\def\IJMPA(#1,#2,#3){Int.\ J.\ Mod. \ Phys.\ A \issue(#1,#2,#3)}
\def\MPLA(#1,#2,#3){Mod.\ Phys.\ Lett.\ A \issue(#1,#2,#3)}
\def\PTP(#1,#2,#3){Prog.\ Theor.\ Phys. \issue(#1,#2,#3)}
\def\RMP(#1,#2,#3){Rev.\ Mod.\ Phys. \issue(#1,#2,#3)}
\def\NIMA(#1,#2,#3){Nucl.\ Instrum.\ Methods \ A \issue(#1,#2,#3)}
\def\EPJC(#1,#2,#3){Eur.\ Phys.\ J.\ C \issue(#1,#2,#3)}
\def\RPP (#1,#2,#3){Rept.\ Prog.\ Phys. \issue(#1,#2,#3)}
\def\PPNP(#1,#2,#3){ Prog.\ Part.\ Nucl.\ Phys. \issue(#1,#2,#3)}
\newc{\PRDR}[3]{{Phys. Rev. D} {\bf #1}, Rapid  Communications, #2 (#3)}

\def\PLB(#1,#2,#3){Phys.\ Lett.\ B  \iss(#1,#2,#3)}
\def\JHEP(#1,#2,#3){JHEP \iss(#1,#2,#3)}

\def\gmin2{(g-2)_\mu}

\catcode`\@=11 

\def\vev#1{\left\langle #1\right\rangle}
\def\lsim{\mathrel{\mathpalette\@versim<}}
\def\gsim{\mathrel{\mathpalette\@versim>}}
\def\@versim#1#2{\vcenter{\offinterlineskip
    \ialign{$\m@th#1\hfil##\hfil$\crcr#2\crcr\sim\crcr } }}
\def\etal{{\em et. al.}}

\def\r2{\sqrt 2}
\def\beq{\begin{equation}}
\def\eeq{\end{equation}}
\def\beqn{\begin{eqnarray}}
\def\eeqn{\end{eqnarray}}
\def\sinW2{\sin^2\theta_W}

\def\mz2{M_{z}^2}
\def\c2b{\cos 2\beta}

\def\m#1{{\tilde m}_#1}

\def\mw#1{{\tilde m}_{\omega #1}}

\def\mz{M_Z}
\def\m0{m_0}
\def\mhalf{m_{\frac{1}{2}}}

\def\cb{\cos\beta}

\def\sec2w{sec^2\theta_W}

\def\gmin2{(g-2)_\mu}

\catcode`\@=11 

\def\vev#1{\left\langle #1\right\rangle}
\def\lsim{\mathrel{\mathpalette\@versim<}}
\def\gsim{\mathrel{\mathpalette\@versim>}}
\def\@versim#1#2{\vcenter{\offinterlineskip
    \ialign{$\m@th#1\hfil##\hfil$\crcr#2\crcr\sim\crcr } }}
\def\etal{{\em et. al.}}

\def\tb{\tilde b}

\def\tL{\tilde L}

\def \chonep{{\wt\chi_1}^{+}}
\def \chonep2{{\wt\chi_2^+}}
\def \chonem2{{\wt\chi_2^-}}

\def \chonepm{{\wt\chi_1}^{\pm}}

\def \mchonepm{m_{\chonepm}}

\def \lstop{\wt{t}_{1}}

\def \lspone{\wt\chi_1^0}
\def \mlspone{m_{\lspone}}
\def \lsptwo{\wt\chi_2^0}

\def\PL{Phys. Lett.}
\def\PRL{Phys. Rev. Lett.}

\def\PR{Phys. Rev.}

\def \lsptwo{\wt\chi_2^0}
\def \lspone{\wt\chi_1^0}
\def \chonem {{\wt\chi_1^-}}
\def \chargino1 {{\wt\chi_1^\pm}}
\def \chargino2 {{\wt\chi_2^\pm}}
\def \lstop{\wt{t}_{1}}
\def \ch2m {{\wt\chi_2^-}}

\def \chonep {{\wt\chi_1^+}}

\def\mygraph#1#2{ \subfigure[]{
   \label{#1}
   \hspace*{-0.6in}
   \begin{minipage}[b]{0.5\textwidth}
   \centering
   \hspace*{4ex}
   \includegraphics[width=\textwidth]{#2}
   \vspace*{-4ex}
   \end{minipage}}
   \vspace*{-1ex}}
\begin{document}
\begin{center}
{\large {\bf {Higgsino Dark Matter in Nonuniversal Gaugino Mass Models}}}
\vglue 0.5cm
Manimala Chakraborti$^{a}$\footnote{tpmc@iacs.res.in},
Utpal Chattopadhyay$^{a}$\footnote{tpuc@iacs.res.in},
Soumya Rao$^{b}$\footnote{soumya.rao@adelaide.edu.au} and
D.P. Roy$^{c}$\footnote{dproy1@gmail.com}\\
{$^a$  Department of Theoretical Physics, Indian Association
for the Cultivation of Science,\\
2A \& B Raja S.C. Mullick Road, Jadavpur,
Kolkata 700 032, India}\\
{$^b$ ARC Centre of Excellence for Particle Physics at the Terascale,  
School of Chemistry \& Physics, 
University of Adelaide, Adelaide, SA 5005, Australia
}\\
{$^c$
Homi Bhabha Centre for Science Education,
Tata Institute of Fundamental Research,
Mumbai-400088, India.
}
\end{center}
\vskip 0.3cm

\begin{abstract}
 We study two simple and well-motivated nonuniversal gaugino mass
models, which predict Higgsino dark matter. One can account for the
observed dark matter relic density along with the observed Higgs boson
mass of $\simeq$~125~GeV over a large region of the parameter space of each
model, corresponding to Higgsino mass of $\simeq$ 1 TeV. In each case this
parameter region covers the gluino mass range of 2-3 TeV, parts of which
can be probed by the 14 TeV LHC experiments. We study these model
predictions for the LHC in brief and for dark matter detection experiments in
greater detail.
\end{abstract}

\newpage
\setcounter{footnote}{0}

\vspace{0.2cm}

\section{Introduction}
\label{sec1}
Supersymmetry, in particular the minimal supersymmetric standard
model (MSSM)\cite{susyreviews1, susyreviews2, susybooks} 
offers a natural candidate for the dark matter\cite{Kamionkowski, Silk} 
of the Universe in the form of the lightest supersymmetric particle (LSP). 
Astrophysical constraints
require it to be a colorless and neutral particle, while direct detection
experiments disfavor a sneutrino dark matter\cite{sneutrino_dd}. 
Thus the favored dark matter (DM) candidate in the MSSM is the lightest neutralino 
$\lspone$ which could be any combination of the neutral gauginos, like the 
bino ($\tilde B$), wino ($\tilde W$) and 
Higgsinos $\tilde H_D$, $\tilde H_U$, i.e.
\begin{equation}
\lspone= N_{11} \widetilde B +N_{12} \widetilde W+N_{13} \widetilde H_D+
N_{14} \widetilde H_U. \nonumber
\label{lspcompositioneqn}
\end{equation}
Here $N_{ij}$ for $i,j=1-4$ refers to elements of the matrix that 
diagonalizes the neutralino mass matrix\cite{susybooks}.

 In the simplest version of this model, called the constrained MSSM (CMSSM) 
or the minimal supergravity (mSUGRA) model\cite{susybooks,msugra_orig}, 
the lightest neutralino as a dark matter candidate\cite{Kamionkowski, Silk} 
is dominantly a bino over most
of the parameter space. Since a bino 
does not carry any gauge charge, its
main annihilation mechanism is the so-called bulk annihilation process via
sfermion exchange. But the Higgs boson mass bound of 114 GeV from LEP\cite{lepsusy}
implied large sfermion masses in this model\cite{uc-dd-ad-sp}, 
which was reinforced now with discovery of the Higgs boson at the LHC 
with a mass of about 125 GeV\cite{higgs_discovery}. 
This implies a
very inefficient bulk annihilation process, resulting in an overabundance of
the dark matter relic density over most of the parameter space. We shall see
below that there are only a few strips of 
parameter space available in the CMSSM  
giving a cosmologically compatible dark matter relic density i.e., the stau
coannihilation, the resonant annihilation, the focus point and the Higgsino
dark matter regions\cite{HBnew,susybooks} - each of which requires some amount of fine-tuning
between SUSY parameters. Moreover, large parts of the stau coannihilation
and the resonant annihilation regions are disfavored by the Higgs boson
mass of about 125 GeV, while most of the 
hyperbolic branch\cite{HB,HBnew}/focus point\cite{FP} 
region is disfavored by
the recent direct dark matter detection experiments\cite{dd_msugra}. While the Higgsino dark
matter region is unaffected by these results, it corresponds to squark and
gluino masses $\gsim$ 8-10~TeV in this model, which cannot be probed at the
LHC\cite{dd_msugra, dmsugra_recent}. Therefore this region has little practical interest at least for LHC
experiments.

In this work we shall study the phenomenology of Higgsino dark matter in
some simple and predictive nonuniversal gaugino mass (NUGM) models based on the
SU(5) grand unified theory (GUT)\cite{etcEllis:1985jn,NathMixedRep,Chattopadhyay:2001mj,                                                            nugminter,ucdphiggsino,nugmvarious,Chattopadhyay:2009fr,
Guchait:2011andothers,Chattopadhyay:2005mv}.
The gaugino mass term in the GUT scale Lagrangian is
bilinear in the gaugino fields, which 
belongs to the adjoint representation of the GUT group.
Thus for the 24-dimensional representation of SU(5) the above 
must transform like one of the representations\footnote{or a linear combination of them} 
occurring in their symmetric product\cite{Chattopadhyay:2009fr}:
\begin{equation}
\label{24_24}
{(24 \times 24)}_{symm} ={\it 1}+{\it 24} + {\it 75}+ \it{200}. 
\label{adjointprod}
\end{equation}

The mSUGRA model considers the singlet representation 
for the gaugino mass term, implying a universal
gaugino mass at the GUT scale. On the other hand, any of the three
nonsinglet representations implies nonuniversal gaugino masses at the same 
scale. Each of these three NUGM models is as predictive
as the CMSSM. We shall see below that the {\it 24} model predicts a 
bino-dominated dark matter, as in the case of the CMSSM. But the {\it 75} and 
the {\it 200} models predict Higgsino-dominated dark matter over the bulk of their
parameter spaces. Thus one can obtain the right amount of dark matter relic 
density by considering a Higgsino mass 
of $\sim$~1~TeV\cite{HBnew,Chattopadhyay:2005mv}.
Unlike the CMSSM, however, this is achieved here naturally with 
a significantly reduced degree of fine-tuning between SUSY 
parameters\cite{nugmfinetuning}. Moreover, for both these NUGM models, 
the cosmologically compatible relic density regions of Higgsino 
dark matter correspond to a gluino mass range of
2-3 TeV, at least a part of which can be probed by the 14 TeV LHC
experiments. Therefore these nonuniversal gaugino mass models should be
of great phenomenological interest in the near future.

In Sec. II we give a brief overview of the above-mentioned universal and
nonuniversal gaugino mass models. In Sec. III we summarize the
phenomenology of the dark matter relic density compatible regions of the
CMSSM. In Sec. IV we describe the dark matter relic density compatible
regions of the {\it 75} model along with the Higgs boson mass constraint. We list
the SUSY mass spectra for a set of benchmark points satisfying these
constraints, which are expected to be within the reach of the 14 TeV LHC
experiments. We also show the size of the gluino pair production cross-section for
these points at the 14 TeV LHC and briefly discusses the signal
characteristics. Then we compare the predictions of this model for various
direct and indirect dark matter detection experiments. In Sec. V we give the
analogous description for the 200 model. We conclude with a summary of
our results in Sec. VI.

\section{Nonuniversality of Gaugino Masses in SU(5) GUT}
\label{sec2}
The gauge kinetic function that relates to the gaugino masses at the GUT
scale originates from the vacuum expectation value of the F-term
of a chiral superfield $\Phi$ which causes SUSY breaking. Thus the gaugino 
masses are obtained via a non-renormalizable dimension-five operator as given 
below\cite{Chattopadhyay:2009fr}:
\begin{equation}
L \supset {\frac{{<F_\Phi>}_{ij}}{M_{Planck}}} \lambda_i \lambda_j .
\label{gauginoLagrangian}
\end{equation}
Here $\lambda_{1,2,3}$ are the U(1), SU(2) and SU(3) gaugino fields (
bino, wino and gluino, respectively).
Since gauginos belong to the adjoint representation of the GUT group, $\Phi$
and $F_\Phi$ can belong to any of the irreducible 
representations occurring in their
symmetric product [Eq. \ref {24_24}], i.e., {\it 1, 24, 75} or {\it 200}. Thus the 
unification scale gaugino
masses for a given representation $n$ of the SUSY-breaking superfield are
determined in terms of one mass parameter $m^n_{1/2}$ as 
\begin{equation}
M^{G}_{1,2,3} = C^n_{1,2,3} m^n_{1/2}, 
\label{relativegauginos}
\end{equation}
 where the values of the coefficients $C^n_{1,2,3}$ are listed in 
Table~\ref{gaugino_mass}\cite{etcEllis:1985jn}.
The coefficients $C^n_3$ are conventionally normalized to 1.

\begin{table}[ht]

\begin{center}
\begin{tabular}[ht]{|l|c|c|c|c|}
\hline
n & $C_3^n$ & $C_2^n$ & $C_1^n$\\
\hline
{\it 1}& 1& 1& 1\\
{\it 24} & 1 & - 3/2 & - 1/2 \\
{\it 75} & 1 & 3 & -5 \\
{\it 200} & 1 &2 & 10\\
\hline
\end{tabular}
\end{center}
\caption{Coefficients $C^n_{1,2,3}$ for the unification scale gaugino mass 
parameters for each representation.}
\label{gaugino_mass} 
\end{table}

The CMSSM assumes the SUSY-breaking superfield $\Phi$ to be a singlet,
implying universal gaugino masses at the GUT scale. On the other hand, any
of the three nonsinglet representations of $\Phi$ would imply nonuniversal
gaugino masses as per Table~\ref{gaugino_mass}. These nonuniversal gaugino 
mass models 
can be consistent with the universality of gauge couplings\footnote{See 
Ref.\cite{Chattopadhyay:2001mj} and references therein.}, 
$\alpha_G \simeq$1/25, and their phenomenology has been widely studied 
\cite{etcEllis:1985jn,NathMixedRep,Chattopadhyay:2001mj,nugminter,ucdphiggsino,
nugmvarious,Chattopadhyay:2009fr,Guchait:2011andothers,Chattopadhyay:2005mv}. 
The superparticle
masses at the electroweak scale are related to these GUT-scale gaugino
masses along with the universal scalar mass parameter $m_0$ 
and trilinear coupling parameter $A_0$, 
via renormalization
 group equations (RGE). 
In particular, the gaugino 
masses evolve
like the corresponding gauge couplings at the one-loop level of the RGE,
implying 
\begin{eqnarray}
M_1 & = & (\alpha_1/\alpha_G) M_1^G  \simeq  (25/60) C_1^n m^n_{1/2}, \nonumber \\
M_2 & = & (\alpha_2/\alpha_G) M_2^G \simeq  (25/30)C_2^n m^n_{1/2}, \nonumber \\ 
M_3 & = & (\alpha_3/\alpha_G) M_3^G  \simeq  (25/9) C_3^n m^n_{1/2}.  
\label{gauginoEW}
\end{eqnarray}
The corresponding Higgsino mass $\mu$ is obtained from the electroweak
symmetry-breaking condition along with the RGE for the Higgs scalar
masses. Neglecting contributions from the trilinear soft terms, 
one has a relatively simple expression for the Higgsino mass at the one-loop level of the RGE\cite{Komine}
, i.e.,
\begin{eqnarray}
\mu^2 + \frac{1}{2} M_Z^2  & \simeq &  
-0.1m_0^2 +2.1 {M_3^G}^2 -0.22 {M_2^G}^2 
-0.006{M_1^G}^2 +0.006 M_1^G M_2^G +  \nonumber \\ 
& & 0.19 M_2^G M_3^G + 0.03 M_1^G M_3^G,
\label{mueqn}
\end{eqnarray}
where the numerical coefficients on the right-hand side correspond to a
representative value of tan$\beta$ = 10, but have only modest variations over the
moderate tan$\beta$ region.

Our results are based on exact numerical solutions of the two-loop RGEs
including also the contributions from the trilinear couplings 
using the S{\sc u}S{\sc pect} code\cite{suspect}. Nevertheless, the
approximate formulae of Eqs.~(\ref{gauginoEW}) and (\ref{mueqn}) are very useful in 
understanding the
composition of the LSP dark matter in these models. The dominant
contribution to the mass of Higgsino (\ref{mueqn}) 
comes from the $M_3^G$ term,
implying $\mu \sim \sqrt 2 m_{1/2}$ from Table~\ref{gaugino_mass} for all  
four models. On the other hand,   
for the mass of the bino, Eq.(\ref{gauginoEW}) 
shows that $M_1 \sim 0.4 m_{1/2}$ for the CMSSM, implying
a bino-dominated LSP dark matter in this model. One sees from Table~\ref{gaugino_mass} that
$M_1$ is further suppressed by a factor of one half in the {\it 24} model, 
implying an even more strongly bino-dominated LSP dark matter. 
Thus one obtains a
generic overabundance of dark matter in the CMSSM as well as in the {\it 24}
model. For the {\it 75} and the {\it 200} models, however, 
one sees from Table~\ref{gaugino_mass} that
the bino mass $M_1$ is enhanced by factors of 5 and 10 ,
respectively, relative to the
CMSSM, implying a Higgsino-dominated LSP dark matter in these
nonuniversal gaugino mass models. Since Higgsino has an efficient
annihilation mechanism via its isospin gauge coupling to the W boson, one obtains a
cosmologically compatible dark matter relic density in these models 
and this corresponds to a Higgsino mass $\mu \simeq$1~TeV\cite{HBnew,Chattopadhyay:2005mv}.

\section{Cosmologically Compatible Dark Matter Relic Density Regions of CMSSM}
\label{sec3}
The cosmologically compatible dark matter relic density regions of the
CMSSM have been thoroughly 
investigated over the last two years in the light
of the 125 GeV Higgs boson mass and other LHC results, as well as by
taking into account the constraints from 
dark matter direct detection experiments.
experiments\cite{dd_msugra,dmsugra_recent}.
We shall briefly revisit this issue here as a
prelude to our investigation of Higgsino dark matter in nonuniversal gaugino
mass models. This will provide a very useful backdrop for comparing the
relative advantage of the dark matter scenario in the latter models. We have
used the S{\sc u}S{\sc pect}\cite{suspect} code in our computation which uses two-loop RGEs 
and radiative electroweak symmetry breaking (REWSB) 
to generate the electroweak scale
SUSY spectra. We consider a  
 theoretical uncertainty of around 3 GeV in the lightest Higgs
scalar mass $m_h$ within the MSSM.  This arises due to the EWSB scale dependence, 
the renormalization scheme (such as the ${\overline{DR}}$ or 
the on-shell schemes as used in S{\sc u}S{\sc pect} and {\sc feynhiggs}\cite{Hahn:2009zz} ,
respectively), uncertainties in the mass of the top quark and higher order loop 
corrections upto 3~loops\cite{higgsuncertainty3GeV}. 
Hence, we assume that a mass of 122 GeV should be consistent with the Higgs 
data.

\begin{figure}[!htb]
\vspace*{-0.05in}
\mygraph{msugra_10}{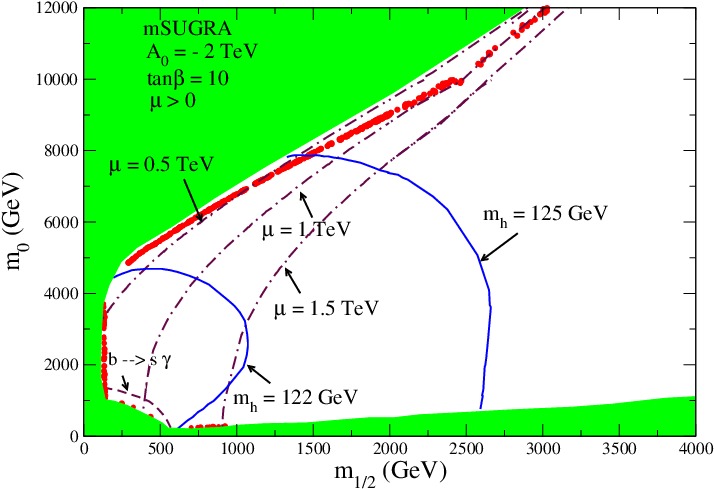}
\hspace*{0.5in}
\mygraph{msugra_50}{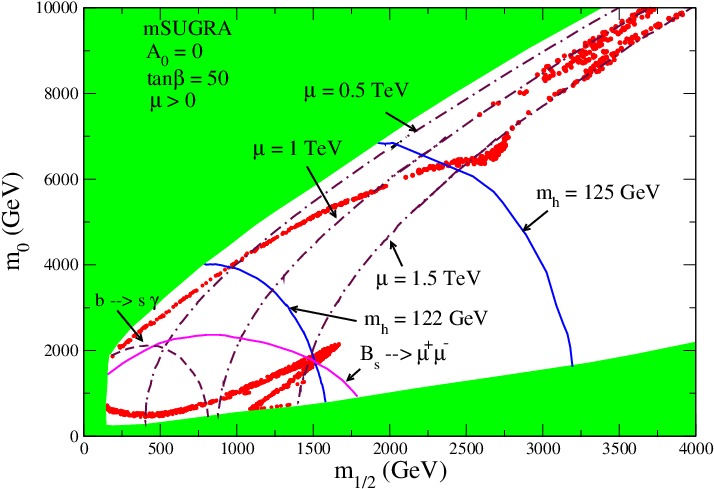}
\caption{The CMSSM/mSUGRA parameter space for representative
values of moderate tan $\beta$ = 10 (a) and large tan $\beta$ = 50 (b). The
cosmologically compatible dark matter relic density regions are indicated by
the red dots, while the constraints from the Higgs boson mass band of 
122-125~GeV are indicated by the blue solid lines. 
The constraints from 
$B_s \rightarrow \mu^+ \mu^-$ 
and $b \rightarrow s \gamma$ 
decays are also indicated by solid magenta and maroon dashed
lines respectively (see text).  The region above these lines is allowed by the corresponding
constraints.
The green region at the top
is mostly excluded due to absence of REWSB ($\mu^2$ turning negative), 
while the green region at the bottom is excluded because of the stau becoming 
the LSP.
}
\label{msugra}
\end{figure}

Figure \ref{msugra} shows the CMSSM parameter space for representative values of
moderate $\tan\beta$ (= 10, Fig.\ref{msugra_10}) and large $\tan\beta$ (= 50, 
Fig.\ref{msugra_50}). 
The shaded 
region on top is disallowed due to lack of REWSB ($\mu^2< 0$), while the bottom strip is
disallowed because the stau becomes the LSP. 
The constraints from the Higgs boson mass
band of 122-125 GeV are indicated by the blue solid lines. Note that one 
requires a fairly large value of the GUT scale trilinear coupling parameter, 
$A_0 = - 2$~TeV, consistent with the charge and color breaking 
constraint\cite{Chattopadhyay:2014gfa} 
to raise the Higgs mass above 122 GeV via the top-squark mixing
contribution at $\tan\beta$ = 10. 
The constraints from the 
$B_s \rightarrow \mu^+ \mu^-$(2$\sigma$)\cite{Aaij:2013aka,Chatrchyan:2013bka,CMSandLHCbCollaborations:2013pla,bsmumuextra} 
and $b \rightarrow s \gamma$(3$\sigma$)\cite{btosgammalimits,bsgammaextra} 
decays are also indicated:
\begin{eqnarray}
2.77 \times 10^{-4} <Br(b \rightarrow s \gamma)<4.09 \times 10^{-4}, \nonumber \\
0.67 \times 10^{-9} <Br(B_s \rightarrow \mu^+ \mu^-) <6.22 \times 10^{-9}.
\label{bphysicslimits}
\end{eqnarray} 
The strips of red dots indicate the
cosmologically compatible dark matter relic density regions, satisfying 
WMAP\cite{wmap}/PLANCK\cite{planck} data,\footnote{The limits correspond to a 
5$\sigma$ range of the PLANCK data that accommodates well the WMAP provided 
range.} 
\begin{equation}
0.09 < \Omega h^2 < 0.14.
\label{omegah2}
\end{equation}  
They are usually classified into the following four regions\footnote{We will 
ignore here the so-called bulk-annihilation region characterized by LSP pair 
annihilation via t-channel slepton exchange 
since it occurs for a smaller $m_{1/2}$ zone that is excluded by 
the Higgs mass data.}.\\
1. The stau coannihilation region is the short strip adjacent to the lower 
boundary in Fig.\ref{msugra_10}, where the LSP dark matter 
co-annihilates with a nearly degenerate stau, 
$\lspone \tilde \tau_1 \rightarrow \tau \gamma (Z)$, via s-channel $\tau$
or t-channel $\tilde \tau$ exchange.  It requires a degeneracy between the bino dark matter and the stau masses to within 10-15 \%.

2. The resonant annihilation region is the funnel-shaped strip in Fig.\ref{msugra_50}, corresponding to s-channel annihilation of the dark matter 
pair into a fermion pair $ \lspone \lspone \rightarrow f \bar f$ 
principally via the pseudoscalar Higgs boson $A$.

Since the $H \lspone \lspone$ and $A \lspone \lspone$ couplings are 
proportional 
to the product of the
gaugino and Higgsino components of $\lspone$ 
the same are strongly
suppressed for a bino-dominated LSP. Therefore it requires the resonance
condition, $M_A \simeq 2M_1$, for enhancement from the Breit-Wigner denominator
along with a large $\tan\beta$ for a large coupling of $A$ 
to the fermion pair. Note
that both the stau coannihilation and resonant annihilation regions require
some fine-tuning between independent SUSY mass parameters. Besides, 
large parts of both regions are disfavored by the Higgs boson mass 
constraint.

3. The hyperbolic branch/focus point region 
near the upper boundary in each part of Fig.\ref{msugra} extends 
upto $m_{1/2} \simeq 3$~TeV.
Here the LSP has a large admixture
of bino and Higgsino components because 
$\mu \sim M_1$. Since the $Z \lspone$ $\lspone$ 
coupling is proportional to the difference of the squares of the 
Higgsino components of $\lspone$ 
(i.e. $N_{13}^2 - N_{14}^2$), the
pair annihilation $\lspone \lspone \rightarrow f \bar f$ 
occurs mainly via Z boson exchange.

Rewriting the electroweak symmetry breaking condition (Eq.\ref{mueqn}) for the CMSSM
in terms of $m_0$ and $m_{1/2}$ one obtains the hyperbolic equation in 
$m_{1/2}$ and $m_0$ for fixed values of $\mu$,
\begin{equation}
\mu^2 + \frac{1}{2} M_Z^2  \simeq  
-0.1m_0^2 + 2 m_{1/2}^2. 
\label{mueqn2}
\end{equation}
\noindent
One can have a 
substantial cancellation between the two terms on the right-hand side,
within the hyperbolic branch/focus point region with $m_0 > > m_{1/2}$.  
This ensures a low value of $\mu \sim M_1 \sim 0.4 m_{1/2}$. 
Note however that it implies a
significant amount of fine-tuning between $m_0$ and $m_{1/2}$. 
Moreover, most of
this region is strongly disfavored by the negative
results from the recent DM direct detection experiments\cite{dd_msugra}. 
The reason is that sizable gaugino and Higgsino components of $\lspone$ in
this region imply a large $H \lspone$ $\lspone$ coupling, predicting a large spin-independent (SI) $\lspone p$ cross-sections for these experiments.

4. Finally, the right end of the strip near the upper boundary corresponds to
$\mu \lsim M_1$, i.e. $\mu \lsim 0.4 m_{1/2}$, implying a Higgsino-dominated 
dark matter in CMSSM\cite{HBnew}.
Since the Higgsino pair can annihilate via their gauge coupling to W bosons,
one obtains the desired dark matter relic density (Eq.\ref{omegah2})
for a Higgsino DM mass $\mu \simeq$~1~TeV, practically independent of any other SUSY parameter. 
The Higgsino DM region is realized for $m_{1/2} \gsim 3$~TeV so that the mass of the bino $M_1 \gsim 1.2$~TeV (Eq.\ref{gauginoEW}) ,
while the mass of corresponding gluino is above 10~TeV. The squark masses are 
also sufficiently heavy, which implies that the strongly interacting 
sparticles are
well beyond the reach of the LHC at 14 TeV. 
The TeV scale superparticle masses can nevertheless easily account for the 
desired Higgs boson mass of $\sim$ 125 GeV. 
However, one sees from the above hyperbolic
equation (Eq. \ref{mueqn2}) that in this case there is at least as 
large a fine-tuning between the
$m_0$ and $m_{1/2}$ parameters as there is in the focus point region. 
We note that the 1~TeV Higgsino signal can be detected via 
the associated single 
photon process
at the 3 TeV CLIC\cite{Chattopadhyay:2005mv,clic}. But it is generally believed that there will be no
CLIC if there is no SUSY signal at the LHC. In that sense this region seems
to be of little practical interest at least for the colliders.

We shall see below that one obtains a Higgsino LSP dark matter with a mass of ~1~TeV in the {\it 75} and {\it 200} models with many properties similar to those of the
CMSSM, but with two major advantages. It occurs naturally in these
nonuniversal gaugino mass models, without requiring any large cancellation
between independent SUSY parameters. Moreover, the corresponding gluino
and top-squark masses lie over the 2-3 TeV region, at least a part of which
are within the reach of 14 TeV LHC. 

\section{Phenomenology of Higgsino Dark Matter in the 75 Model}
\label{sec4}
\begin{figure}[!htb]
\vspace*{-0.05in}
\mygraph{75_10}{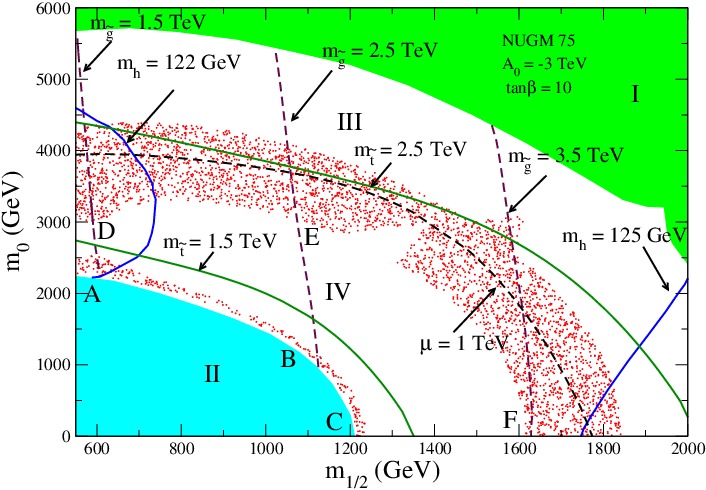}
\hspace*{0.5in}
\mygraph{75_30}{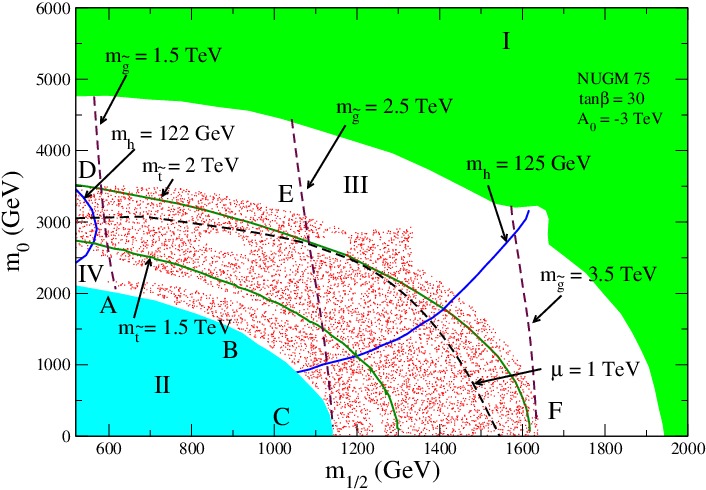}
\caption{(a) Plot in the $m_0-m_{1/2}$
plane for the {\it 75} 
model for tan$\beta=$10.  Region I is excluded because of a nonconvergent EWSB 
solution. Region II
is disfavored as the top-squark becomes the LSP or tachyonic there.
Contours for top-squark, gluino masses, $\mu = 1$ TeV, $m_h=125$ GeV and 
$m_h=122$ GeV are also shown.
Red points satisfy the relic density constraint (Eq.~\ref{omegah2}).
For the red points lying along the boundary of region II 
in the strip ABC, the LSP is Higgsino like.
Along the strip DEF also we find the LSP to be Higgsino like.
In the region ABC the main DM annihilating mechanisms are
coannihilations involving $\lstop$, $\chonepm$, $\lspone$ and $\lsptwo$.
All the way along the strip DEF coannihilations occur where
$\chonepm$, $\lspone$ and $\lsptwo$ take part almost
equally.  In regions III and IV we obtain underabundant and overabundant DM respectively.
The entire parameter space is allowed by $B_s \rightarrow \mu^+ \mu^-$ and $b \rightarrow s \gamma$
constraints.
(b)Similar plot as in panel (a) with tan$\beta=$30. Colours and conventions are the same as plot in panel (a).
Here we get underabundant DM for both the regions III and IV.}
\label{75_m0_mhalf}
\end{figure}

Fig.\ref{75_m0_mhalf} shows the $m_{1/2}-m_0$ parameter plane of 
the NUGM 
model corresponding to the representation {\it 75} of SU(5) GUT for
representative values of $\tan\beta$ = 10 and 30 when $A_0=-3$~TeV. 
Region I at the top is 
excluded due to the non-convergent EWSB solution, while region II at the bottom is excluded
due to the lighter top-squark ($\tilde t_1$) being the LSP/tachyonic for both 
Fig.\ref{75_10} and Fig.\ref{75_30}. 
The regions with red dots correspond to higgsino dark matter that satisfy the cosmological relic density, 
while the constraints from the Higgs boson mass range of 122-125 GeV are
indicated by the blue solid lines. Contours of gluino and lighter stop masses 
are indicated along with the $\mu$ = 1 TeV contour. In the band DEF the
cosmological relic density of dark matter is achieved through coannihilation
among the degenerate charged and neutral Higgsinos ($\lspone, \lsptwo, \chonepm$ ), while in the
strip ABC near the lower boundary there is additional coannihilation with
the lighter top-squark ($\tilde t_1$).
 In regions III and IV we obtain underabundant and overabundant 
DM, respectively, for Fig.\ref{75_10}, whereas the regions labeled    
III and IV correspond to only underabundant DM in Fig.\ref{75_30}. 
$B_s \rightarrow \mu^+ \mu^-$ and $b \rightarrow s \gamma$ constraints are
satisfied everywhere.
We now comment on the 
appearance of the {\it clip} regions in both 
of the above figures. We have investigated it by varying $m_0$ for a given $m_{1/2}$ 
in relevant regions. 
A jump in $\mu$ appears near a particular zone of $m_0$ associated with 
the {\it clip} region. 
This is essentially associated with the way the corrections 
to $\mu^2$ arising out of a finite order effective potential  
(one or two loop) are computed\cite{Arnowitt:1992qp}. 
It is possible that a logarithmic 
term for the correction (typically from the top-squark contribution) 
may turn from a negative value to a positive value almost 
discontinuously for a small change in $m_0$ around a given value of 
$m_0$. This would give rise a jump in the value of $\mu$. 
For a Higgsino-dominated LSP, such an abrupt,  
albeit small change in $\mu$ may mean a significant amount of change in the 
relic density ($\sim \mu^2$). Thus the above explains the appearance of 
{\it clip} regions within the zone that satisfies relic density. 
A similar effect occurs in the zone near the REWSB boundary.   
Of course, the inclusion of higher order 
terms in the effective potential would smooth out such jumps or eliminate 
the {\it clip} regions in general.  

%

Table~\ref{75_bmp} lists the superparticle masses for three benchmark 
points (BP) from the
left part of each figure namely Figs. \ref{75_10} and \ref{75_30} 
with relatively light gluinos, which can be probed at
the high luminosity run of the 14 TeV LHC. Here we show the gluino
pair-production cross-section for these points at the 14 TeV LHC 
as obtained by using the code  
{\sc prospino}\cite{prospino}. These cross-sections correspond to several hundred 
gluino pairs at 
100~${\rm fb}^{-1}$ which can be probed in the high luminosity 
run of LHC. 
Moreover, the probe can be extended up to a gluino mass of 2.5~TeV
 at the very high luminosity runs of 1000-3000 
${\rm fb}^{-1}$\cite{Baer:2012vr}.
We note that Table~\ref{75_bmp} 
shows an inverted
hierarchy of squark masses with a relatively light top-squark ${\tilde t}_1$. Together with the
large coupling of the top-squark with the Higgsino, it implies that the gluino will
dominantly decay via a real or virtual top-squark: 
$\tilde g \rightarrow t \bar t \tilde \chi_{1,2}^0$.
In the second to last row we show the branching ratio (BR) results from 
{\sc susy-hit}\cite{Djouadi:2006bz} for the dominant decay modes of $\gl$. Only the 
modes having BR $>$~10\% are shown.
For BPs 1, 3, 4 and 6, $\gl$ decays to $\tilde t_1 \bar t$ type of 
final states with 100 \% BR.  
As shown in the last row, $\tilde t_1$ further decays to $b \chonep$ since kinematically there is no 
way to have a top and an LSP for the above decay.
On the other hand, for BPs 2 and 5, the $\gl \ra \tilde t_1 \bar t$ decay mode is kinematically 
forbidden leading to $\gl$ decaying into a three body final states.
It is clear from the BRs that the signal will contain two to four top quarks 
plus a large $\met$ from the decay of the gluino pair. 
Moreover, the Majorana nature of the gluino implies that 
half of the two top quarks plus $\met$ final states 
will have a same sign top quark pair. Thus, one expects a 
distinctive signal with either two same sign top quarks or 
three to four top quarks accompanied by a large $\met$ 
from the LSP pair. Depending on the BPs or the parameter space in general 
one needs to carefully compute the SM backgrounds\cite{SM3and4tops}. 
We hope the members 
of the ATLAS and CMS collaborations will make detailed simulation studies of this signal, which is beyond the scope of the present work.
We shall proceed now to the model predictions for the direct and indirect
dark matter detection experiments. 

\begin{table}[ht]
{\tiny
\begin{center}
\begin{tabular}[ht]{|l||c|c|c||c|c|c|}
\hline
Parameter & 1 & 2 & 3 & 4 & 5 & 6\\
\hline
\hline
$m_{1/2}$ &          837.11 &      731.80 &      658.32 &      843.10 &      762.16 &      656.10   \\
$m_0$ &             1948.54 &     3689.09 &     2248.34 &     1805.52 &     3110.81 &     2089.04   \\
$\tan\beta$ &         10.00 &       10.00 &       10.00 &       30.00 &       30.00 &       30.00   \\
$A_0$&     -3000.00 &    -3000.00 &    -3000.00 &    -3000.00 &    -3000.00 &    -3000.00   \\
$(M_1,M_2,M_3)$ &
1884, 2035. 1767  &     1661, 1786, 1517 & 1475,1599,1400 &  1895,2052,1786  & 1724, 1860, 1593 &  1467,1595, 1403\\
$\mu$ &             1268.56 &     1062.37 &     1235.64 &     1176.45 &      978.18 &     1148.12   \\
$m_{\tilde g}$ &            1954.41 &     1822.47 &     1600.97 &     1957.89 &     1861.66 &     1587.32   \\
$m_{{\tilde \chi_1}^{\pm}}, m_{{\tilde \chi_2}^{\pm}}$ &            1272.51,    2081.25 &     1072.96,    1854.31 &     1234.37,    1650.76 &     11\
81.44,    2095.35 &      988.12,    1922.20 &     1148.78,    1643.44   \\
$m_{{\tilde \chi_1}^0}, m_{{\tilde \chi_2}^0}$ &            1272.18,    1275.75 &     1072.59,    1076.29 &     1233.99,    1240.02 &     1180.92,  \
  1183.69 &      987.55,     990.32 &     1148.18,    1152.51   \\
$m_{\tilde t_1}, m_{\tilde t_2}$ &          1298.52,    2421.22 &     2188.85,    3365.72 &     1282.69,    2287.72 &     1270.36,    2235.20 &     \
1909.76,    2838.59 &     1220.38,    2054.05   \\
$m_{\tilde b_1}, m_{\tilde b_2}$ &          2412.53,    2486.19 &     3364.03,    3866.90 &     2279.98,    2533.74 &     2098.72,    2229.44 &     \
2835.95,    3048.80 &     2038.31,    2127.47   \\
$m_{\tilde u_1}, m_{\tilde u_2}$ &          2922.94,    2671.38 &     4104.89,    3982.31 &     2818.42,    2661.21 &     2848.60,    2584.39 &     \
3657.64,    3500.96 &     2697.29,    2531.82   \\
$m_{\tilde e_L}, m_{\tilde e_R}$&           2620.75,    2480.56 &     3978.62,    3919.23 &     2633.97,    2550.39 &     2527.42,    2377.81 &     \
3485.48,    3405.96 &     2499.53,    2410.04   \\
$m_{{\tilde \tau}_1}, m_{{\tilde \tau}_2}$ &        2450.19,    2606.89 &     3881.97,    3960.64 &     2519.42,    2619.58 &     2096.37,    2402.7\
6 &     3078.22,    3331.77 &     2122.94,    2368.87   \\
$m_A, m_{H^{\pm}}$ &        2873.09,    2873.11 &     4068.20,    4068.24 &     2865.53,    2865.33 &     2210.31,    2210.25 &     2913.68,    2913\
.65 &     2179.37,    2179.33   \\
$m_h$ &              123.13 &      122.04 &      122.20 &      123.95 &      122.87 &      123.03   \\
$\Omega_{\tilde \chi_1}h^2$ &          0.11 &        0.13 &        0.12 &        0.12 &        0.11 &        0.10   \\
$BF(b\to s\gamma)$ &  3$\times 10^{-4}$ & 3.04$\times 10^{-4}$ & 3$\times 10^{-4}$ & 2.77$\times 10^{-4}$ & 2.91$\times 10^{-4}$ &
2.75$\times 10^{-4}$ \\
$BF(B_s\to \mu^+\mu^-)$ & 3.53$\times 10^{-9}$ & 3.53$\times 10^{-9}$ & 3.53$\times 10^{-9}$ &3.76$\times 10^{-9}$ & 3.58$\times 10^{-9}$ &
3.76$\times 10^{-9}$ \\
$\sigma_{p\chi}^{SI}$ in pb & 7$\times 10^{-10}$ & 7.57$\times 10^{-10}$ & 2.97$\times 10^{-9}$ & 4.54$\times 10^{-10}$ & 4.11$\times 10^{-10}$ &
1.74$\times 10^{-9}$\\
$\sigma_{gg}^{NLO}$ in fb & 1.23 & 2.83 & 8.45 & 1.19 &
2.21 & 8.97\\
\hline
\hline
  &  &  &  &  &  &  \\
Dominant  & $\gl \ra \wt t_1 \bar t$ 50 & $ \gl \ra \lspone t \bar t $ 23  &
$\gl \ra \wt t_1 \bar t $  50 & $\gl \ra \wt t_1 \bar t $ 50 & $\gl \ra \lspone t \bar t $ 23 &  $\gl \ra \wt t_1 \bar t$ 50 \\
decay   & $\quad \ra \wt t_1^{*} t$ 50 & $\quad \ra \lsptwo t \bar t $ 18 & $\quad \ra \wt t_1^{*} t $  50  &
$\quad \ra \wt t_1^{*} t $  50  &$\quad \ra \lsptwo t \bar t $ 20  & $\quad \ra \wt t_1^{*} t $  50 \\
modes of $\tilde g$ (in \%)&    & $\quad  \ra \chonem t \bar b $ 27 & &  & $\quad  \ra \chonem t \bar b $ 27 & \\
($>$ 10 \% are shown)    &   & $\quad  \ra \chonep b \bar t $ 27 &  & & $\quad  \ra \chonep b \bar t $ 27 & \\

\hline
  &  &  &  &  &  &  \\
Dominant decay & $\tilde t_1 \ra b \chonep$ 100 & -  & $\tilde t_1 \ra b \chonep$  100 & $\tilde t_1 \ra b \chonep$ 100 & - & $\tilde t_1 \ra b \chonep$ 100\\
modes of $\tilde t_1$ (in \%) &  & & & &  &  \\
($>$ 10 \% are shown)    &  & & & &  &  \\
\hline

\end{tabular}
\end{center}
\caption{Spectra of six benchmark points for the 75 model. Masses and 
mass parameters are
shown in GeV. Gluino pair production cross sections correspond to a 14 TeV LHC run. The relevant SM parameters used are $m_t^{pole}= 173.5$~GeV,
$m_b^{\overline{MS}} = 4.18$~GeV and $m_{\tau}=1.77$~GeV.  Branching ratios for the dominant decay 
modes of $\gl$ and ${\tilde t}_1$ are also shown.
}
\label{75_bmp}
}
\end{table}
\clearpage

    The spin-independent scattering cross section 
$\sigma^{\rm SI}_{p \lspone}$ 
of the nucleon with $\lspone$ involves Higgs exchange (t-channel) or 
squark exchange (s-channel) 
diagrams. 
With the present LHC limit of squark masses, the Higgs exchange 
processes dominate in $\sigma^{\rm SI}_{p \lspone}$ . 
Typically, unless the LSP is a mixture of 
Higgsino with bino or wino the couplings are suppressed.   
For a Higgsino-dominated scenario of LSP with 
$\tilde{\chi}_1^0$ ($|\mu| << M_1, M_2$) and for the 
decoupling limit of the Higgs boson ($M_Z^2<<M_A^2$) one finds 
the relevant couplings \cite{Hisano:2004pv} 
$C_{h \tilde \chi \tilde \chi}$ and $C_{H \tilde \chi \tilde \chi}$ 
that explicitly show the suppression effect when $M_1$ and $M_2$ are 
away from $\mu$. The results are not however valid  
when $|\mu|$ is close to either $M_1$ or $M_2$.  
\begin{eqnarray}
C_{h \tilde \chi \tilde \chi} & \simeq & \mp \frac{1}{2} M_Z c_W
\bigl[ 1 \pm \sin2\beta \bigr]
    \biggl[ \frac{t_W^2}{M_1 - |\mu|}  + \frac{1}{M_2 - |\mu|} \biggr], \nonumber \\
C_{H \tilde \chi \tilde \chi} & \simeq & \frac{1}{2} M_Zc_W \cos2\beta
    \biggl[ \frac{t_W^2}{M_1 - |\mu|}  + \frac{1}{M_2 - |\mu|} \biggr] ,
\label{higgsinoSIcouplings}
\end{eqnarray}
for $\mu>0$ and $\mu<0$, respectively, with $s_W=\sin{\theta_W}$ etc. The 
particular result to note from the above equation for 
the Higgsino-dominated LSP case is that the direct detection SI 
cross-section decreases with increase in gaugino masses $M_1$ and $M_2$.  
Thus for a given $m_{1/2}$, 
the {\it 75} model will have a decreased value for 
$\sigma^{\rm SI}_{p \lspone}$ because of larger values of the 
bino and wino masses when compared with the mSUGRA scenario. 
Figs. \ref{tanb10_75_si} and \ref{tanb30_75_si} show the results of 
$\sigma^{\rm SI}_{p \lspone}$ for different values of the LSP mass
for $\tan\beta=10$ and $\tan\beta=30$, 
using micrOMEGAs\cite{Belanger:2013oya}. 
This corresponds to the parameter 
space of Fig.\ref{75_m0_mhalf}.  
The exclusion contours from 
XENON100\cite{xenon100} and LUX\cite{lux} are also shown in addition to the 
estimated exclusion level for future XENON1T\cite{xenon1t} experiment. The regions satisfying the relic density 
bracketed within $900< \mlspone< 1300$~GeV for $\tan\beta=10$ 
and $900< \mlspone< 1200$~GeV for $\tan\beta=30$ are shown in red and the 
points generally satisfy the LUX limit. On the other hand a large region 
of parameter space may be probed via the future XENON1T experiment.

\begin{figure}[!htb]
\mygraph{tanb10_75_si}{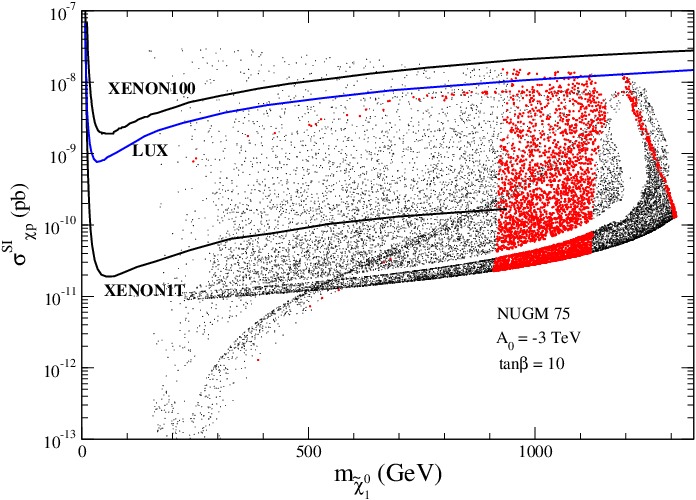}
\hspace*{0.5in}
\mygraph{tanb30_75_si}{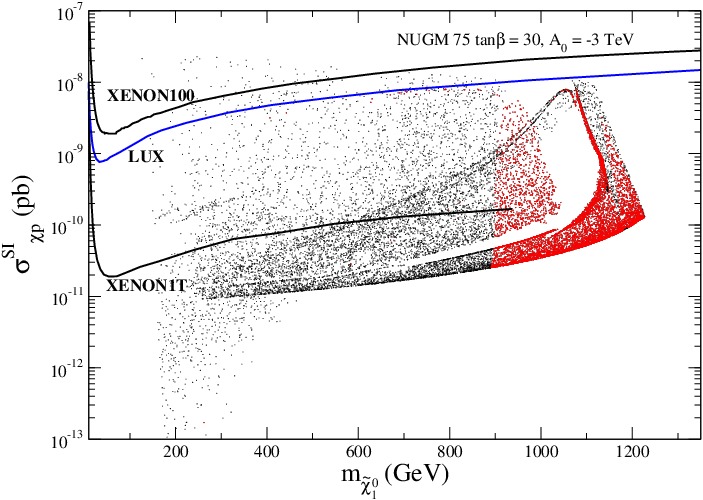}
\caption{ (a) spin-independent scattering cross-section of the LSP with a proton
as a function of LSP mass for the {\it 75} model with tan$\beta=$10.
The constraints coming
from direct detection experiments like XENON100,
LUX and the expected limit from XENON1T are shown. The points in red satisfy the relic
density constraint.
(b) Similar plot as in panel
(a) for tan$\beta=$30.}
\label{75_si}
\end{figure}

Coming to the spin-dependent (SD) LSP-proton cross section 
we note that $\sigma_{p\chi}^{SD}$ is associated with a 
Z exchange since the large values of the squark masses after the LHC data 
would not cause any significant contribution from squark exchange diagrams. 
The coupling of $\lspone \lspone Z$ related to a Higgsino asymmetry 
is given by $C_{Z \tilde \chi \tilde \chi}={|N_{13}^2 - N_{14}^2|}$.  
For a Higgsino-like LSP one can have the following approximate expression 
\cite{Hisano:2004pv, Barger:2008qd}:
\begin{equation} 
C_{Z \tilde \chi \tilde \chi} \simeq \mp 
\frac{1}{2} 
    \biggl[ {t_W^2} \frac{M_W^2}{M_1 \mu}  + \frac{M_W^2}{M_2 \mu} \biggr] 
\cos 2\beta +{\cal O} (\frac{\mu}{M_1}, \frac{\mu}{M_2}),
\label{higgsinoSDcouplings}
\end{equation}   
for $\mu>0$ and $\mu<0$ respectively. 
We note that for the region of
parameter space satisfying the relic density, a Higgsino-dominated LSP when associated with sufficiently 
large electroweak gaugino masses (which is indeed true for both the 
NUGM models considered in this work) results in a significant amount 
of suppression of $\sigma_{p\chi}^{SD}$. This is visible in 
Fig.\ref{tanb10_75_sd} as well as in Fig.\ref{tanb30_75_sd},  
where we show the scatter plots of $\sigma_{p\chi}^{SD}$ vs $\mlspone$ 
for $\tan\beta=10$ and 30 
,respectively, for the {\it 75} model. 
For the 
zones of $\mlspone$ satisfying the relic density, $\sigma_{p\chi}^{SD}$ (as shown with red dots) 
is way too small to be 
probed via the shown IceCube exclusion limits (both the existing and the 
projected limits). Here, the spin-dependent cross section is  
obtained via indirect means by searching for muon neutrinos at 
IceCube\cite{Icecube:2012} arising out of dark matter annihilation within the Sun. 
We will also discuss the muon flux limit in relation to the mass of dark 
matter in this section. 
We may mention that in the present scenario, the IceCube limits 
are stronger \cite{IceCube:2011aj, Cushman:2013}
than the dedicated spin-dependent 
direct detection experiments like COUPP\cite{coupp}.

\begin{figure}[!htb]
\mygraph{tanb10_75_sd}{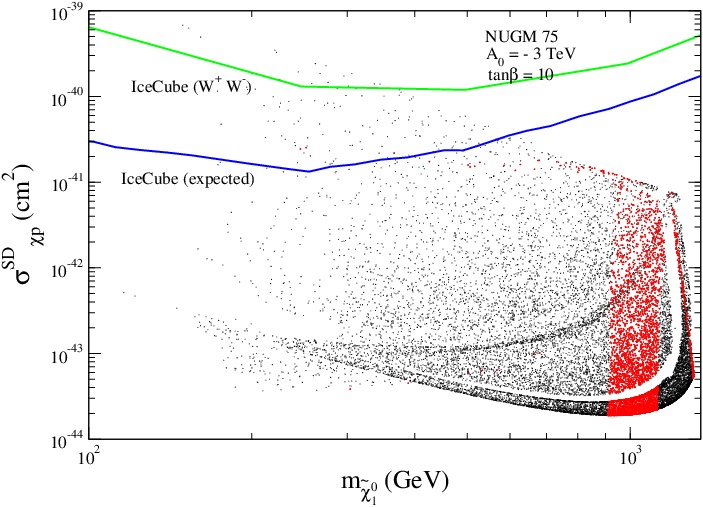}
\hspace*{0.5in}
\mygraph{tanb30_75_sd}{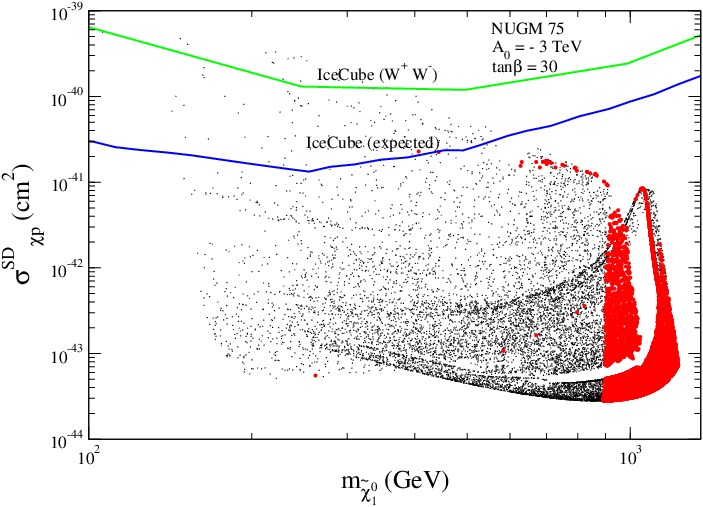}
\caption{(a) Variation of the spin-dependent cross section with the LSP mass
  for the {\it 75} model with tan$\beta=$10.
The IceCube exclusion limit for the $\lspone \lspone \rightarrow W^+W^-$ channel is
shown as a green line \cite{IceCube:2011aj}.  The blue line represents the expected sensitivity reach of IceCube.
(b) Similar plot as in panel (a) for tan$\beta=$30.}
\label{75_sd}
\end{figure}

\vspace{0.20cm}

   We next consider DM indirect detection studies for the {\it 75} model for 
the photon signal. 
There can be a sufficient degree of gravitational capture of 
weakly interacting massive particles (WIMPs) due to nuclear scattering 
effects. Gravitational capture may occur in the dense regions ,
like galactic centers, dwarf galaxies or in the core regions of 
objects within the 
Solar System such as the Sun 
or the Earth\cite{Kamionkowski,Silk}. The LSP pair-annihilation 
would produce fermion-antifermion pairs or electroweak 
gauge bosons. Decays of products of primary annihilation 
and hadronization may produce $\pi^0$ that would eventually produce 
photons. Apart from the above
there can be final state radiation effects (FSR) of primarily produced 
particles. We note that the environment of gravitational capture and 
LSP annihilation is associated with a much smaller velocity ($v/c \sim 10^{-3} 
$) unlike a much larger velocity existing at the time of freeze-out.
Thus, the annihilation of LSPs in the present day scenario involves 
a large $p$-wave suppression [${(v/c)}^2$]. 
We remind ourselves that the LSP being a Majorana particle the combined CP property and the combined parity of the LSP pair are the same.
This makes the favored 
s-channel particle (namely, the CP-odd Higgs boson $A$) contribute dominantly 
to the photon signal, which on the other hand is p-wave suppressed 
as discussed above. We note that a larger Higgsino content is generally 
favorable for the photon signal. 
However, the NUGM models under discussion depend on high scale input 
parameters and involve RGEs and REWSB that lead to correlated SUSY spectra.  
All these cause an s-channel Higgs resonance to become a remote possibility.   
Figs.\ref{tanb10_75_sigmav} and \ref{tanb30_75_sigmav} show 
$<\sigma v>$, the thermally averaged LSP-pair annihilation cross-section
 as a function of $\mlspone$ for the
{\it 75} model with $\tan\beta=$10 and $\tan\beta=$30 respectively. 
The Fermi-LAT constraint from LAT dwarf spheroidal stacking (4 years)\cite{fermi-lat-gamma} 
is shown as a green line.  The red points correspond to 
parameter points satisfying the relic density. Owing to a larger $b \bar b$ coupling of 
the Higgs for a large $\tan\beta$, $<\sigma v>$ is generally larger 
in Fig.\ref{tanb30_75_sigmav} in comparison with Fig.\ref{tanb10_75_sigmav}. 
The parameter space for the 
{\it 75} model is practically unconstrained by the present Fermi limit.

\begin{figure}[!htb]
\mygraph{tanb10_75_sigmav}{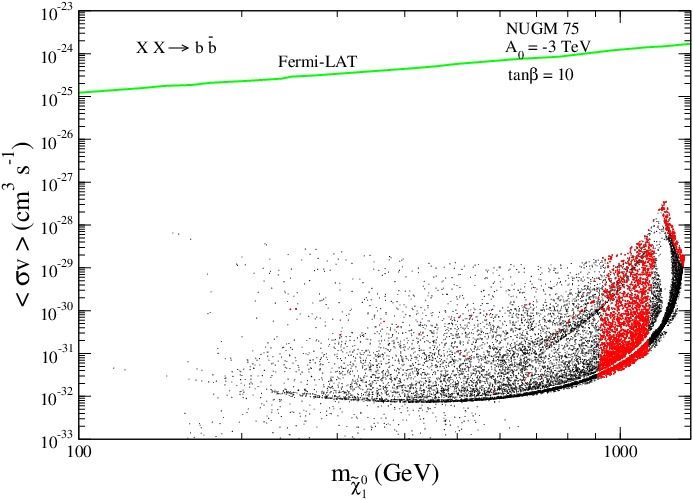}
\hspace*{0.5in}
\mygraph{tanb30_75_sigmav}{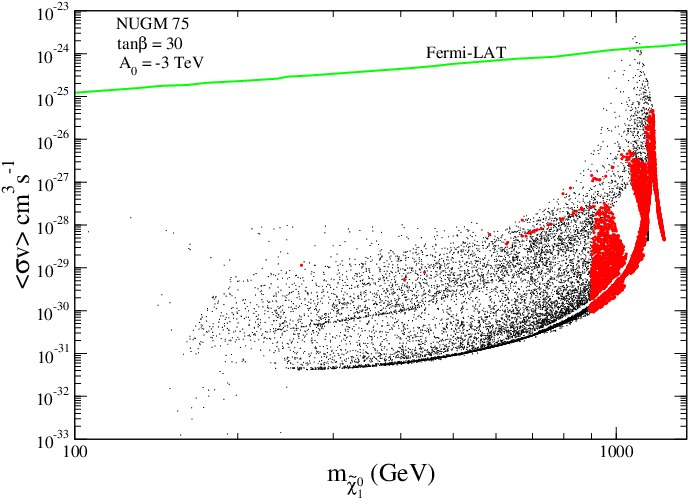}
\caption{ (a) DM self-annihilation cross-section as a function of DM mass for the
{\it 75} model with tan$\beta=$10.
Fermi-LAT constraint [LAT dwarf spheroidal stacking (4 years)]
\cite{fermi-lat-gamma}
is shown as a green line.
The parameter space is practically
unconstrained by Fermi data. (b) Similar plot as in panel
(a) for tan$\beta=$30.}
\label{75_sigmav}
\end{figure}

We would like to discuss the probing of Higgsino-dominated 
dark matter via indirect detection of muon flux at IceCube\cite{Icecube:2012} 
due to neutrinos from the Sun. 
In SUSY models neutrinos cannot be produced at tree level 
in neutralino annihilations. However, neutrinos may arise from other sources,
like heavy quarks, gauge bosons, tau leptons etc\footnote{There is a
possibility of having neutrinos from
two-to-two annihilation into gauge bosons via loops\cite{Ibarra:2014}
}. Thus neutrinos 
are produced with a broad energy distribution, with energy 
reaching up to a sizable 
fraction of the DM mass. For a DM mass less than $M_W$, neutrinos from 
$b \bar b$ or $\tau^+ \tau^-$ are the primary channels. But these are 
not very promising candidates for detection with given experimental 
thresholds of neutrino detection. 
For massive neutralinos, annihilations may additionally 
lead to gauge bosons, top quarks or Higgs bosons. A neutralino with a  
substantial Higgsino component may undergo pair annihilation to produce 
gauge bosons which in turn may produce high energy neutrinos.
We must keep in mind that neutrinos of energy several hundreds of GeV 
produced inside the Sun would be depleted since the probability of a 
neutrino to escape the Sun without interaction is given by 
$P=e^{-E_\nu/E_k}$, where depending on the type of neutrino 
$E_k$ varies from 130 to 230 GeV\cite{Silk}.
 Neutrino 
oscillation is taken into account while computing the flux of muon 
neutrinos at the detector. At the detector, the muon flux 
arising from neutrinos via charge-current interactions is detected . 
    
Neutrino signals from the Sun or other dense regions of galaxy in general 
involve capture and annihilation of WIMPs. 
In general both spin-independent and spin-dependent 
types of scattering of WIMPs with various nuclei 
may lead to appreciable reduction of energy leading to the WIMP velocity 
going below the escape velocity.  This leads to WIMPs  
being captured within the object and 
also undergoing pair annihilations. Thus the time evolution of $N$ WIMPs is
\begin{equation}
\frac{dN}{dt}= C-C_A N^2.
\label{wimpdepletion}
\end{equation}  
Here $C$ refers to the rate at which WIMPs are captured
 and $C_A$ 
depends on the annihilation cross section of WIMPs and is related to the WIMP 
annihilation rate $\Gamma_A$  via $\Gamma_A=\frac{1}{2}C_A N^2$ in the Sun 
\cite{Silk, Gould, Griest:1986}
Any possibility to have a positive evaporation term that is linear in $N$ is 
neglected here. The above arises out of a scenario of WIMP-nuclear scattering 
where the WIMP is much lighter than a given nucleus in abundance in the Sun. 
Such terms may potentially increase the speed of WIMPs
above the escape velocity\cite{Buckley:2013}.   
The time-dependence of $N$ from Eq.\ref{wimpdepletion} leads to 
$\Gamma_A \equiv \frac{1}{2}C_A N^2=\frac{1}{2}C \tanh^2(t/\tau)$ where 
$\tau=1/\sqrt{C C_A}$. With appreciably large capture 
and annihilation rates that indeed is possible  
for the Sun for various models including supersymmetry and with the present 
time $t=t^\odot=4.5 \times 10^9$~years,   
it is realistic to assume $t/\tau >>1$  leading to $\Gamma_A=\frac{1}{2}C$.  
This of course
means an equilibrium scenario out of capture and annihilation of WIMPs\cite{Wikstrom:2009}.
This is
however not true for capture and annihilation of WIMPs in the Earth which is much less 
massive leading to either a much smaller escape velocity or 
the dominance of spin-independent interactions in the WIMP-nuclear 
scattering resulting in reduced capture rates for WIMPs. 
Thus probing DM via muon flux due to neutrino propagation 
is not so promising for the Earth when compared to the prospect for the Sun\cite{Silk}. 
We must note that both 
SI and SD cross-sections are important for capture of WIMPs in the 
Sun\cite{Chacko, Ibarra:2014}.
Capture cross sections may be related through suitable models to SI and SD 
WIMP-nuclear cross sections and 
it is through such relations that the measurement of muon flux due to 
neutrino signals may 
be translated into setting limits on SI and SD cross sections\cite{Silk, Ibarra:2014}.  
Figs. \ref{tanb10_75_mu_flux} and \ref{tanb30_75_mu_flux} show the 
results of muon flux with respect to the mass of LSP for $\tan\beta=10$ 
and 30, respectively, for the {\it 75} model. The 
IceCube exclusion limit for the $\lspone \lspone \rightarrow W^+W^-$ 
channel is shown as a green line\cite{IceCube:2011aj}.  
The blue line represents the expected sensitivity reach of IceCube. 
Clearly, IceCube would not be able to probe the region of parameter 
space that satisfies the relic density limits. 

\begin{figure}[!htb]
\mygraph{tanb10_75_mu_flux}{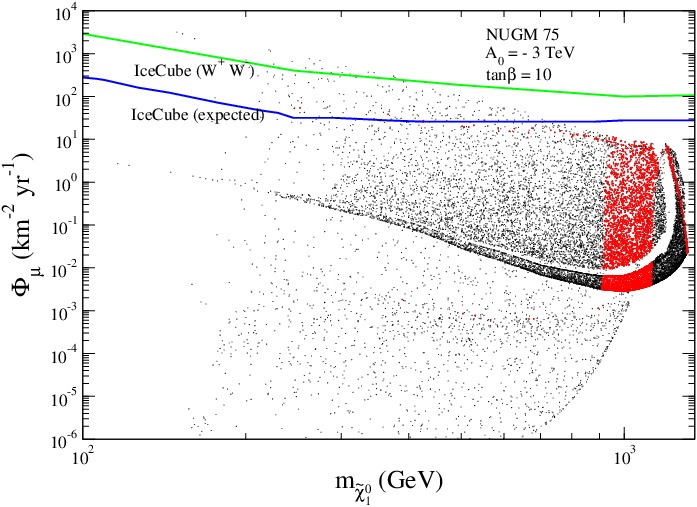}
\hspace*{0.5in}
\mygraph{tanb30_75_mu_flux}{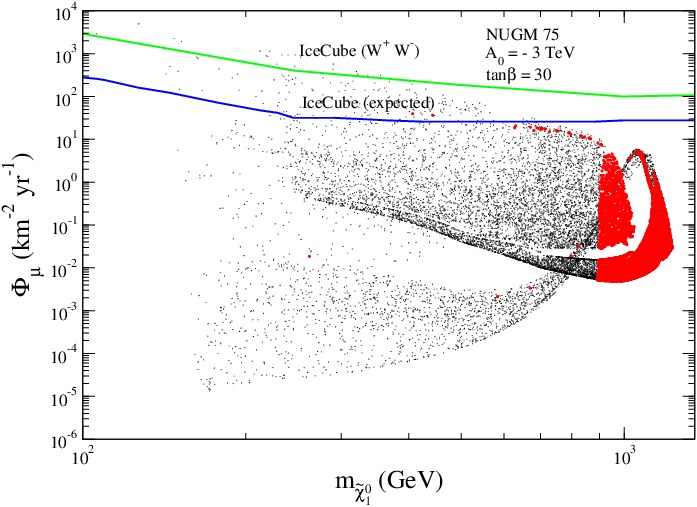}
\caption{(a) Variation of the muon flux with the LSP mass
for the {\it 75} model with tan$\beta=$10.
The IceCube exclusion limit for the $\lspone \lspone \rightarrow W^+W^-$ channel is
shown as a green line\cite{IceCube:2011aj}.  The blue line represents the expected sensitivity of IceCube.
(b) Similar plot as in panel (a) for tan$\beta=$30.}
\label{75_mu_flux}
\end{figure}

\section{Phenomenology of Higgsino Dark Matter in the 200 Model}
\label{sec5}
Figs. \ref{tanb10_200_m0_mhalf} and \ref{tanb30_200_m0_mhalf} show the 
scatter plots in the $\mhalf-m_0$ plane of 
the NUGM model corresponding to the representation {\it 200} of SU(5) GUT for
representative values of $\tan\beta$ = 10 and 30 when $A_0=-2$~TeV. 
Region I is excluded because of nonconvergent EWSB solution. Region II
is disallowed because lighter top-squark (${\tilde t}_1$) 
becomes the LSP or tachyonic. 
Contours for squark and gluino 
masses for a few different values along with the contours for 
$\mu=$ 1 TeV, $m_h=122$ GeV and $m_h=125$~GeV are also shown.
Red points satisfy the relic density constraint.  
In the region A the LSP is Higgsino-like with very little wino admixture.
The mechanisms that allow the DM to satisfy the relic density constraint are
coannihilation processes among $\chonepm$, $\lspone$ and $\lsptwo$.
Along the branches B and C the LSP is also Higgsino like.
There are coannihilations involving $\chonepm$, $\lspone$ and $\lsptwo$. 
Here additionally, we
find $\mlspone \simeq \mchonepm \simeq M_A/2$.
Thus we find s-channel Higgs ($A,H, H^\pm$)
resonance processes involving coannihilations
among $\lspone$ and/or $\chonepm / \lsptwo$.
Along the strips DE and EF we also get mostly a
Higgsino like LSP. Here coannihilation processes involving 
$\chonepm$, $\lsptwo$ and $\lspone$
cause the dark matter to achieve
the right relic density.                                                       
For regions III, IV and V we get underabundant DM, whereas regions VI and 
VII give overabundant DM.  
The entire parameter space respects $B_s \rightarrow \mu^+ \mu^-$ and $b \rightarrow s \gamma$
constraints.

\vspace{0.08cm}

\noindent
Table~\ref{200_bmp} shows the superpartner masses and other data of 
phenomenological interest 
for three benchmark points each for $\tan\beta=10$ and 30 corresponding to 
Figs. \ref{tanb10_200_m0_mhalf} and \ref{tanb30_200_m0_mhalf}, respectively.
The mass patterns are more or less not very different from 
the {\it 75} model. However, we must keep in mind that $M_1^G$ is significantly larger (by a factor of 2) whereas $M_2^G$ is smaller (by a factor of 
$\frac{2}{3}$) for the {\it 200} model when compared with the {\it 75} model 
(see Table~\ref{gaugino_mass}). 
Consequently, the masses of left and right components of scalars are 
affected differently via RGE effects.   
The change happens such that 
almost all the squarks and the sleptons are more split among the left 
and the right scalars in the {\it 200} model.  
The top-squark sector has a reduced splitting because of smaller $|A_0|$ considered 
here compared to the {\it 75} model.
The second to last row shows the gluino pair-production cross-section [at  
next-to-leading order (NLO)]
at the 14 TeV LHC using {\sc prospino}\cite{prospino}. Typically these would
correspond to fewer gluino pairs compared to the {\it 75} model. 
Nonetheless they correspond to around 100 events in the high 
luminosity 100~${\rm fb}^{-1}$ run of LHC.
The decay modes of $\gl$ are more or less the same as those of the {\it 75} model for BPs 1, 2 and 5 leading to similar signal characteristics. 
For BPs 3, 4 and 6 there is an additional 
decay mode : $\gl \ra \tilde b_1 \bar b$ which was absent 
in the  {\it 75} case.  
The reason lies in the fact that $\tilde t_1$ and $\tilde b_1$ have 
similar masses here compared to the {\it 75} case. 
Furthermore, the last row shows the relevant BRs for the decays of $\tilde t_1$ and $\tilde b_1$. We note that $\wt b_1$ dominantly decays in the channel 
$\wt b_1 \ra t \chonem$. 
So as in the {\it 75} model, the gluino pair will decay dominantly into two, three or four top quarks plus $\met$, with half of the two top events having same sign tops. 

\begin{figure}[!htb]
\mygraph{tanb10_200_m0_mhalf}{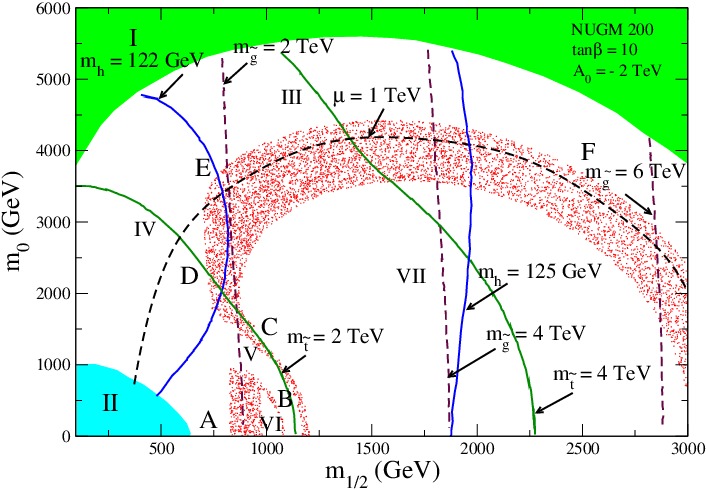}
\hspace*{0.5in}
\mygraph{tanb30_200_m0_mhalf}{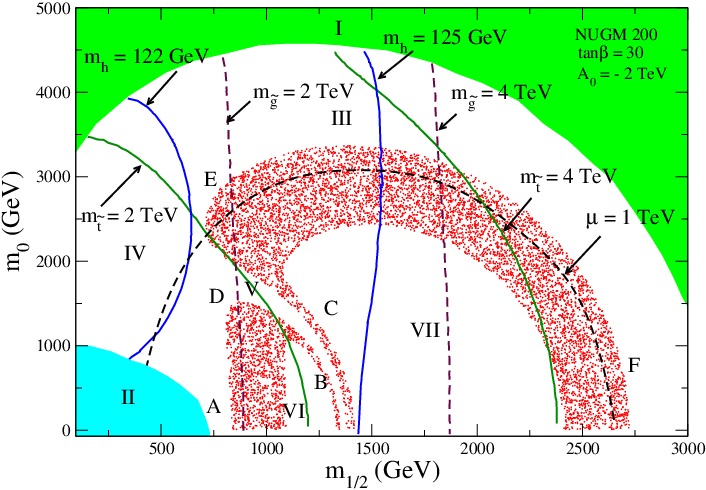}
\caption{ (a) Allowed parameter space in the $m_0-m_{1/2}$
plane for the {\it 200} model for tan$\beta=$10.  Region I is excluded because of a
nonconvergent EWSB solution. Region II
is disallowed because top-squark becomes the LSP or tachyonic.  
Contours for the squark, gluino
masses, $\mu=$ 1~TeV, $m_h=122$~GeV and $m_h=125$~GeV are also shown.
Red points satisfy the relic density
constraint.  In region A the LSP is Higgsino-like with very little wino admixture.
The mechanisms that allow the DM to satisfy WMAP relic density constraint are
coannihilation processes among $\chonepm$, $\lspone$, $\lsptwo$.
Along the branches B and C the LSP is Higgsino like.
There are coannihilations involving $\chonepm$, $\lspone$ and $\lsptwo$.
Here, additionally we also
find $\mlspone \simeq \mchonepm \simeq M_A/2$.
Thus we find s-channel Higgs ($A,H, H^\pm$)
resonance processes involving coannihilations
among $\lspone$ and/or $\chonepm / \lsptwo$.
Along the strips DE and EF we also get mostly a
Higgsino like LSP and coannihilation among $\chonepm$, $\lsptwo$ and  $\lspone$
helps the dark matter to achieve
the right relic density.
For regions III, IV and V we get underabundant DM, whereas regions VI and VII give overabundant DM.
The parameter space is unconstrained by $B_s \rightarrow \mu^+ \mu^-$ and $b \rightarrow s \gamma$
limits.
(b) Similar plot as in panel
(a) for tan$\beta=$30. }
\label{200_m0_mhalf}
\end{figure}

\noindent
\begin{table}[ht]
{\tiny
\begin{center}
\begin{tabular}[ht]{|l|c|c|c|c|c|c|}
\hline
Parameter & 1 & 2 & 3 & 4 & 5 & 6 \\
\hline

\hline
$m_{1/2}$ &          848.94 &      818.46 &      874.26 &      833.33 &      732.93 &      881.86   \\
$m_0$ &             1663.05 &     2847.18 &      895.64 &     1102.62 &     2587.57 &      644.69  \\
$\tan\beta$ &         10.00 &       10.00 &       10.00 &       30.00 &       30.00 &       30.00   \\
$A_0$&     -2000.00 &    -2000.00 &    -2000.00 &    -2000.00 &    -2000.00 &    -2000.00   \\
$(M_1,M_2,M_3)$ &
3781,1364,1779 & 3670,1318,1695 & 3880,1402,1842 & 3696,1338,1763 & 3272,1181,1534 & 3912,1416, 1865 \\
$\mu$ &             1224.92 &     1096.58 &     1278.14 &     1180.55 &      962.14 &     1224.61   \\
$m_{\tilde g}$ &            1976.53 &     1969.82 &     1989.92 &     1911.12 &     1779.53 &     1993.02   \\
$m_{{\tilde \chi_1}^{\pm}}, m_{{\tilde \chi_2}^{\pm}}$ &            1215.42,    1420.40 &     1096.80,    1380.77 &     1264.42,    1453.96 &     11\
71.41,    1386.67 &      961.65,    1238.37 &     1216.89,    1459.56   \\
$m_{{\tilde \chi_1}^0}, m_{{\tilde \chi_2}^0}$ &            1214.18,    1235.18 &     1095.45,    1110.36 &     1263.25,    1286.26 &     1170.02,  \
  1187.81 &      960.01,     974.02 &     1215.54,    1231.19   \\
$m_{\tilde t_1}, m_{\tilde t_2}$ &          1935.09,    2143.44 &     2413.91,    2711.43 &     1660.69,    2004.42 &     1577.30,    1957.53 &     \
2151.43,    2334.72 &     1512.69,    1995.50   \\
$m_{\tilde b_1}, m_{\tilde b_2}$ &          2002.13,    2467.81 &     2696.88,    3302.78 &     1699.20,    2111.78 &     1605.97,    1946.00 &     \
2298.94,    2754.09 &     1533.65,    1863.84   \\
$m_{\tilde u_1}, m_{\tilde u_2}$ &          2545.05,    3064.03 &     3362.44,    3751.19 &     2203.44,    2812.92 &     2213.15,    2771.82 &     \
3049.81,    3394.37 &     2137.80,    2770.35   \\
$m_{\tilde e_L}, m_{\tilde e_R}$&           2510.23,    3534.60 &     3364.87,    4134.98 &     2140.52,    3338.13 &     2156.49,    3258.35 &     \
3047.09,    3731.57 &     2066.61,    3308.62   \\
$m_{{\tilde \tau}_1}, m_{{\tilde \tau}_2}$ &        2493.40,    3510.79 &     3346.80,    4105.56 &     2123.61,    3316.60 &     2001.68,    3055.4\
0 &     2888.06,    3469.38 &     1907.25,    3112.53   \\
$m_A, m_{H^{\pm}}$ &        2757.87,    2757.70 &     3498.83,    3498.88 &     2462.24,    2462.44 &     2011.60,    2011.60 &     2612.90,    2612\
.90 &     1966.87,    1966.85   \\
$m_h$ &              122.42 &      122.01 &      122.90 &      123.42 &      122.45 &      123.77   \\
$\Omega_{\tilde \chi_1}h^2$ &          0.11 &        0.12 &        0.10 &        0.10 &        0.09 &        0.10   \\
$BF(b\to s\gamma)$ & 3.02$\times 10^{-4}$ & 3.04$\times 10^{-4}$ & 3.01$\times 10^{-4}$ & 2.78$\times 10^{-4}$ & 2.88$\times 10^{-4}$ &
2.78$\times 10^{-4}$  \\
$BF(B_s\to \mu^+\mu^-)$ & 3.53$\times 10^{-9}$ & 3.53$\times 10^{-9}$ & 3.54$\times 10^{-9}$ & 3.78$\times 10^{-9}$ & 3.59$\times 10^{-9}$ & 3.81$\times 10^{-9}$ \\

$\sigma_{p\chi}^{SI}$ in pb & 1.41$\times 10^{-8}$ & 7.46$\times 10^{-9}$ & 1.65$\times 10^{-8}$ & 1.15$\times 10^{-8}$ & 6.54$\times 10^{-9}$ & 9.44$\times 10^{-9}$ \\


$\sigma_{gg}^{NLO}$ in fb & 1.09 & 1.26 & 9.73$\times 10^{-1}$  &
1.47 & 3.31 & 9.51$\times10^{-1}$ \\
\hline
\hline
  &  &  &  &  &  &  \\
Dominant  & $\gl \ra \lspone t \bar t$ 12 & $\gl \ra \lspone t \bar t $ 20  &
$\gl \ra \wt b_1 \bar b $  29.5 & $\gl \ra \wt b_1 \bar b $ 29 & $\gl \ra \lspone t \bar t $ 17 &  $\gl \ra \wt b_1 \bar b $ 26  \\
decay   & $\quad \ra \lsptwo t \bar t $  11 & $\quad \ra \lsptwo t \bar t $ 17 & $\quad \ra \wt b_1^{*} b $  29.5 &
$\quad \ra \wt b_1^{*} b $  29 &$\quad \ra \lsptwo t \bar t $ 14  & $\quad \ra \wt b_1^{*} b $  26\\
modes of $\tilde g$ in (\%) & $\quad  \ra \chonem t \bar b $ 22 & $\quad  \ra \chonem t \bar b $ 23 & $\quad  \ra \wt t_1 \bar t$ 20.5
 & $\quad \ra \wt t_1 \bar t$ 21  & $\quad  \ra \chonem t \bar b $ 23  & $\quad \ra \wt t_1 \bar t$ 22 \\
($>$ 10 \% are shown) & $\quad  \ra \chonep b \bar t $ 22 & $\quad  \ra \chonep b \bar t $ 23 & $\quad  \ra \wt t_1^{*} t$ 20.5 &
 $\quad  \ra \wt t_1^{*} t$ 21 & $\quad  \ra \chonep b \bar t $ 23 &  $\quad  \ra \wt t_1^{*} t$ 22\\
\hline
          &    &   &   &    &   &   \\
Dominant                      &   &  & $\wt t_1 \ra t \lspone$ 24       & $\wt t_1 \ra t \lspone$ 24   &   &$\wt t_1 \ra t \lspone$ 23    \\
decay modes                   &    &    & $\quad \ra t \lsptwo$ 38     & $\quad \ra t \lsptwo$  36   &   & $\quad \ra t \lsptwo$ 42   \\
of $\wt t_1 / \wt b_1$ (in \%)&-   &-   &  $\quad  \ra b \chonep$ 27     & $\quad  \ra b \chonep$ 30   &-   &$\quad  \ra b \chonep$ 33 \\
($>$ 10 \% are shown)         &   &    &   $\wt b_1 \ra t \chonem$ 64   & $\wt b_1 \ra t \chonem$ 61   &   &$\wt b_1 \ra t \chonem$ 74   \\
                              &   &    &  $\quad  \ra t \chtwom$ 23    &  $\quad  \ra t \chtwom$ 14  &   &  $\quad  \ra b \lspone$ 14  \\
                              &   &    &                                & $\quad  \ra b \lspone$ 10   &   & $\quad  \ra b \lsptwo$ 10  \\
\hline
\end{tabular}
\end{center}
\caption{Spectra of six benchmark points for the 200 model. Masses and mass parameters are
shown in GeV. Gluino pair production cross sections correspond to 
a 14 TeV LHC run. Branching ratios for the dominant decay modes of $\gl$, $\tilde t_1$ 
and $\tilde b_1$ are also shown.} 
\label{200_bmp}
}
\end{table}

Figs. \ref{tanb10_200_si} and \ref{tanb30_200_si} show the 
scatter plots of the spin-independent direct detection cross section of 
neutralino dark matter with respect to the mass of LSP in the {\it 200} model 
for $\tan\beta=10$ and 30 respectively.  
The exclusion contours from
XENON100\cite{xenon100} and LUX\cite{lux} are also shown in addition to the
region $900< \mlspone< 1400$~GeV (which satisfies relic density) for both values of $\tan\beta$
are shown in red and an appreciable number of 
parameter points are discarded via the LUX limit. On the other hand a large region
of parameter space may be probed via the future XENON1T experiment.
We note that 
the {\it 200} model has a smaller wino mass $(M_2)$ compared to that for the 
{\it 75} model for a given value of $m_{1/2}$. Thus 
the SI cross-section can be understood to be larger for the {\it 200} model 
(Eq.\ref{higgsinoSIcouplings}) due to the relative closeness of values 
between $\mu$ and $M_2$. Consequently, a larger region of parameter space in the 
{\it 200} model is excluded via the LUX limit in comparison with the 
{\it 75} model (Fig.\ref{75_si}).

\begin{figure}[!htb]
\mygraph{tanb10_200_si}{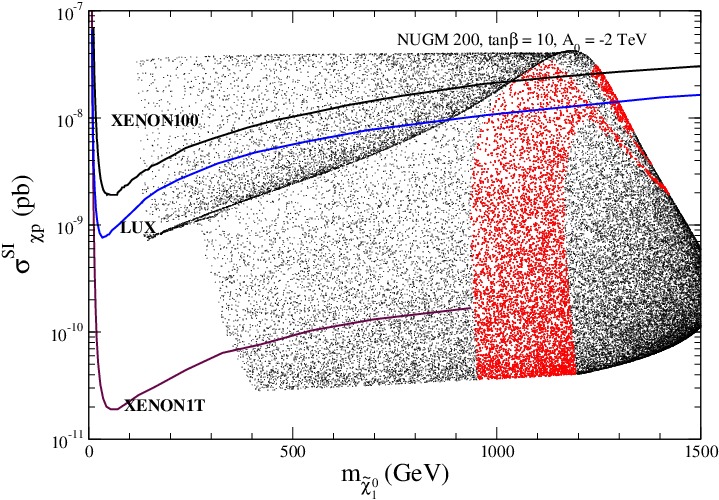}
\hspace*{0.5in}
\mygraph{tanb30_200_si}{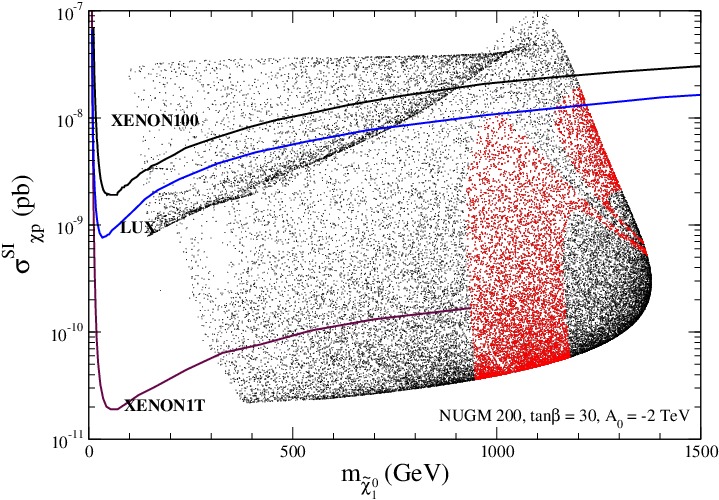}
\caption{ (a) spin-independent scattering cross-section of the LSP with a proton
as a function of LSP mass for the {\it 200} model with tan$\beta=$10.  
The constraints coming
from direct detection experiments like XENON100,
LUX and the future XENON1T are shown. (b) Similar plot as in panel
(a) for tan$\beta=$30.}
\label{200_si}
\end{figure}

Figs. \ref{tanb10_200_sd} and \ref{tanb30_200_sd} show  
the scatter plots of $\sigma_{p\chi}^{SD}$ vs $\mlspone$ 
for $\tan\beta=10$ and 30,
respectively, for the {\it 200} model. 
For the 
zones of $\mlspone$ satisfying the relic density, $\sigma_{p\chi}^{SD}$ (as shown with red dots) 
is way too small to be 
probed via the shown IceCube exclusion limits (both the existing and 
projected limits). As mentioned before, here the spin-dependent cross section is  
obtained via indirect means by searching for muon neutrinos at 
IceCube\cite{Icecube:2012} arising out of dark matter annihilation within 
the Sun. 
We note that in comparison with the {\it 75} model (Fig.\ref{75_sd}) the SD-cross section 
is a little larger in the {\it 200} model because of the relatively smaller 
mass of the wino (Eq.\ref{higgsinoSDcouplings}). 
We will soon discuss the muon flux limit in relation to the mass of dark 
matter.

\begin{figure}[!htb]
\mygraph{tanb10_200_sd}{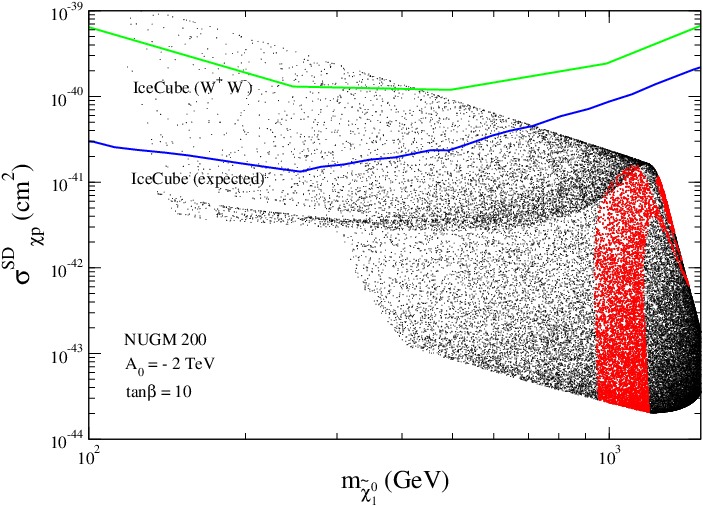}
\hspace*{0.5in}
\mygraph{tanb30_200_sd}{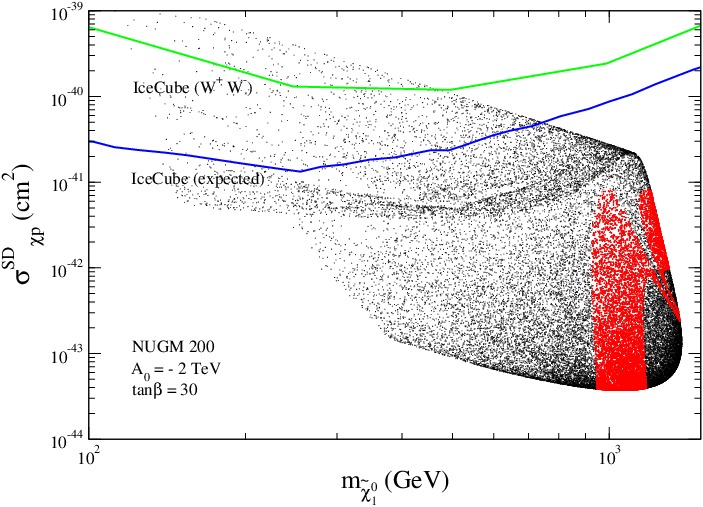}
\caption{(a) Variation of spin-dependent cross section with the LSP mass
  for the {\it 200} model with tan$\beta=$10.
The IceCube exclusion limit for the $\lspone \lspone \rightarrow W^+W^-$ channel is
shown as a green line.  The blue line represents the expected sensitivity reach of IceCube.
(b) Similar plot as in panel (a) for tan$\beta=$30.}
\label{200_sd}
\end{figure}

Figs. \ref{tanb10_200_sigmav} and \ref{tanb30_200_sigmav} show 
$<\sigma v>$ as a function of $\mlspone$ for the
{\it 200} model with $\tan\beta=$10 and $\tan\beta=$30, respectively. 
The Fermi-LAT constraint (LAT dwarf spheroidal stacking (4 years)\cite{fermi-lat-gamma})
is shown as a green line.  The red points correspond to parameter points
that satisfy the relic density bound. 
Similar to Fig.\ref{75_sigmav},  the parameter space for the 
{\it 200} model is practically unconstrained by the present Fermi limit.

\begin{figure}[!htb]
\mygraph{tanb10_200_sigmav}{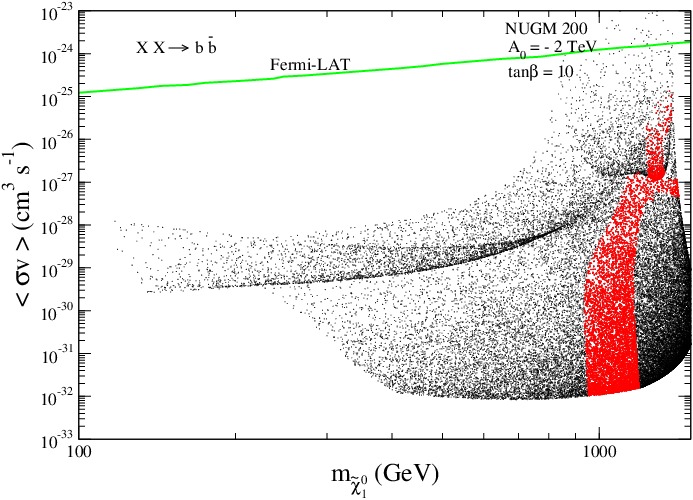}
\hspace*{0.5in}
\mygraph{tanb30_200_sigmav}{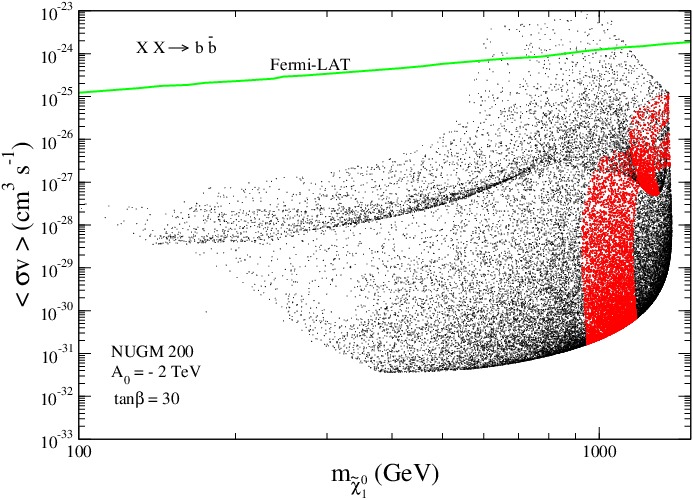}
\caption{ (a) DM self annihilation cross-section as a function of DM mass for the
{\it 200} model with tan$\beta=$10.
The Fermi-LAT constraint [LAT dwarf spheroidal stacking (4 years)]
\cite{fermi-lat-gamma} is shown as a green line.
The parameter space is practically
unconstrained by Fermi data. (b) Similar plot as in panel
(a) for tan$\beta=$30.}
\label{200_sigmav}
\end{figure}



Figs. \ref{tanb10_200_mu_flux} and \ref{tanb30_200_mu_flux} show the 
results of muon flux with respect to the mass of LSP for $\tan\beta=10$ 
and 30, respectively, for the {\it 200} model. The 
IceCube exclusion limit for the $\lspone \lspone \rightarrow W^+W^-$ 
channel is shown as a green line\cite{IceCube:2011aj}.  
The blue line represents the expected sensitivity reach of IceCube. 
Clearly, IceCube would not be able to probe the region of parameter 
space that satisfies the relic density limits. 

\begin{figure}[!htb]
\mygraph{tanb10_200_mu_flux}{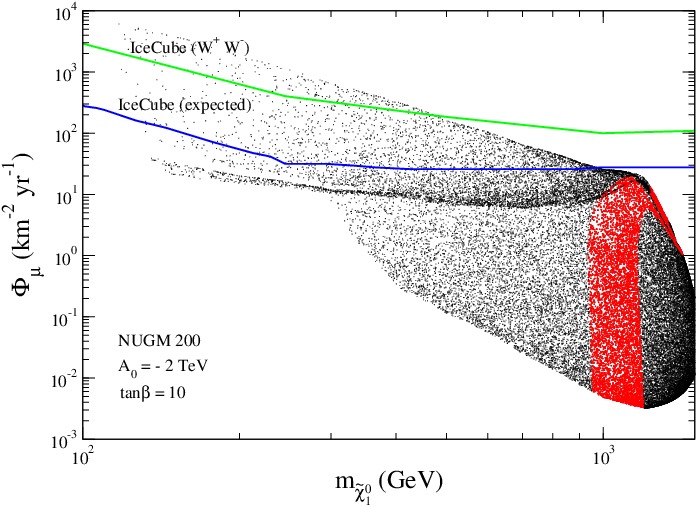}
\hspace*{0.5in}
\mygraph{tanb30_200_mu_flux}{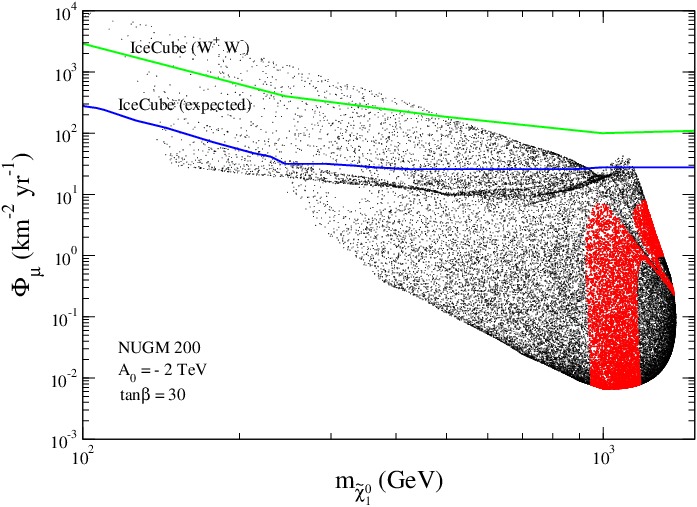}
\caption{(a) Variation of the muon flux with the LSP mass
for the {\it 200} model with tan$\beta=$10.
The IceCube exclusion limit for the $\lspone \lspone \rightarrow W^+W^-$ channel is
shown as a green line.  The blue line represents the expected sensitivity reach of IceCube.
(b) Similar plot as in panel (a) for tan$\beta=$30.}
\label{200_mu_flux}
\end{figure}

\clearpage
\section{Conclusion}
The LHC SUSY searches and discovery of a Higgs boson at 125 GeV have put strong lower bounds on superparticle masses. Consequently, the CMSSM/mSUGRA, with a typically bino-dominated LSP leads to an overabundance of DM relic density over most of its parameter space. There are only a few strips of parameter space that give a relic density compatible with  WMAP/PLANCK data, each of which requires a significant amount of fine-tuning amongst the SUSY mass parameters. Moreover, large parts of the stau coannihilation region and the resonant annihilation region are disallowed by the Higgs mass constraint, while the focus point region is strongly disfavored by the direct DM search experiments.
The Higgsino LSP region can account for the right DM relic density for an LSP mass of about 1 TeV, while satisfying the Higgs mass and other experimental constraints; however, it implies large squark/gluino masses $\gsim$~8-10~TeV, which are inaccessible at LHC. On the other hand, nonuniversal gaugino mass models corresponding to the 75 and 200 representations of SU(5) GUT group naturally predict  a Higgsino-dominated LSP, which can account for the right DM relic density for an LSP mass of about 1 TeV as in the case of the CMSSM but with
much reduced fine-tuning. Moreover, it implies gluino masses in the region of 2-3 TeV in these models, at least a part of which is accessible to high luminosity LHC runs at 14 TeV. We listed the SUSY spectra for a set of benchmark points in this region of the two nonuniversal gluino mass models along with the corresponding gluino pair-production cross-sections at 14 TeV LHC. We also briefly discussed the distinctive signatures of these signal events. We then discussed the prospects of detecting these two model signals in various direct and indirect DM detection experiments. For both of the models these signal cross-sections turn out to be quite small. 
The smallness of the spin-independent direct detection cross-section 
$\sigma_{p\chi}^{SI}$ in the above two models arises from the fact that 
i) the LSP is mostly Higgsino-like with very little bino or wino components and 
ii) the masses of the bino and wino in the two 
models are large for a given mass of gluino in comparison to what is found 
in the CMSSM. 
$\sigma_{p\chi}^{SI}$
is a little higher in the {\it 200} model compared to that
in the {\it 75} model because of a relatively smaller wino mass
for the former model.  
The results show that a significant amount of parameter space is 
allowed by LUX and will 
 be probed by future direct detection experiments like XENON1T.  
We also evaluated the spin-dependent cross-section $\sigma_{p\chi}^{SD}$ for 
the two models. It was found that for the characteristic zones of $\mlspone$ 
that satisfy the relic density limits, the masses of the bino and wino are 
sufficiently high so as to cause some suppression effect. $\sigma_{p\chi}^{SD}$ 
becomes quite small to be probed 
via IceCube.  
Regarding the indirect detection signals, the photon signal intensity is 
small because of a general lack of s-channel Higgs resonance arising out of the 
characteristic spectra of NUGM models that involve the given mass relations among 
gaugino mass parameters, RGEs and REWSB. The 
thermally averaged annihilation cross-section lies well below the Fermi-LAT 
limit.  Similarly the muon flux values are too low to be probed by 
IceCube. One finds that the two NUGM models would be probed better using the
measurement of the spin-independent direct detection cross section via XENON1T 
rather than any other direct and indirect detection of dark matter 
experiments.

\section{Acknowledgment}
D.P.R was partly supported by the senior scientist fellowship of Indian National Science Academy. M.C. would like to thank the Council of 
Scientific and Industrial Research, Government of India for support.



\clearpage

\begin{thebibliography}{99}
\markright{Bibliography}
\bibitem{susyreviews1}
For reviews on supersymmetry, see, {\it e.g.},
H. P. Nilles, \PREP (110, 1, 1984);
J.~D.~Lykken, [\href{http://arxiv.org/abs/hep-th/9612114}{arXiv:hep-th/9612114}];
J. Wess and J. Bagger, {\it Supersymmetry and Supergravity}, 2nd ed., (Princeton, 1991).
\bibitem{susyreviews2}
H. E. Haber and G. Kane, \PREP (117, 75, 1985);
S. P. Martin, [\href{http://arxiv.org/abs/hep-ph/9709356}{arXiv:hep-ph/9709356}];
D. J. H. Chung {\it et al.},  Phys.\ Rept.\  {\bf 407}, 1 (2005) [\href{http://arxiv.org/abs/hep-ph/0312378}{arXiv:hep-ph/0312378}].
 
\bibitem{susybooks}
M. Drees, P. Roy and R. M. Godbole, {\it Theory and Phenomenology of Sparticles}, (World Scientific, Singapore, 2005);
H. Baer and X. Tata, {\it{Weak scale supersymmetry: From superfields to scattering events}}, Cambridge, UK: Univ. Pr. (2006) 537 p.

\bibitem{Kamionkowski} 
C.~Jungman, M.~Kamionkowski and K.~Griest, \PREP (267,195,1996);

\bibitem{Silk}
G.~Bertone, D.~Hooper and J.~Silk, Phys.\ Rept.\  {\bf 405}, 279 (2005).

\bibitem{sneutrino_dd}
  T.~Falk, K.~A.~Olive and M.~Srednicki,
  Phys.\ Lett.\ B {\bf 339}, 248 (1994)
  [hep-ph/9409270].
\bibitem{msugra_orig}
A.~H.~Chamseddine, R.~Arnowitt and P.~Nath,
Phys.\ Rev.\ Lett.\  {\bf 49}, 970 (1982);
R.~Barbieri, S.~Ferrara and C.~A.~Savoy,
Phys.\ Lett.\ B {\bf 119}, 343 (1982);
L.~J.~Hall, J.~Lykken and S.~Weinberg,
Phys.\ Rev.\ D {\bf 27}, 2359 (1983);
P.~Nath, R.~Arnowitt and A.~H.~Chamseddine,
Nucl.\ Phys.\ B {\bf 227}, 121 (1983);
N. Ohta, Prog. Theor. Phys. {\bf 70}, 542 (1983);\\
For a review see: P. Nath, R. Arnowitt and A.H. Chamseddine,
{\it Applied N =1 Supergravity} (World Scientific, Singapore, 1984)
                         

\bibitem{lepsusy}The LEP SUSY Working Group, 
[\href{http://lepsusy.web.cern.ch/lepsusy/}{http://lepsusy.web.cern.ch/lepsusy/}].

\bibitem{uc-dd-ad-sp} 
  U.~Chattopadhyay, D.~Das, A.~Datta and S.~Poddar,
  Phys.\ Rev.\ D {\bf 76}, 055008 (2007)
  [arXiv:0705.0921 [hep-ph]].

\bibitem{higgs_discovery} ATLAS Collaboration, \PLB(716,1-29,2012)  
CMS Collaboration, \PLB(716,30-61,2012).  

\bibitem{HBnew}
U.~Chattopadhyay, A.~Corsetti and P.~Nath,
  Phys.\ Rev.\  D {\bf 68}, 035005 (2003)
  [arXiv:hep-ph/0303201]; 
 S.~Akula, M.~Liu, P.~Nath and G.~Peim,
  Phys.\ Lett.\ B {\bf 709}, 192 (2012)
  [arXiv:1111.4589 [hep-ph]].

\bibitem{HB}
K.~L.~Chan, U.~Chattopadhyay and P.~Nath,
  Phys.\ Rev.\  D {\bf 58}, 096004 (1998)
  [arXiv:hep-ph/9710473].
\bibitem{FP}
J.~L.~Feng, K.~T.~Matchev and T.~Moroi,
Phys.\ Rev.\ D {\bf 61}, 075005 (2000);
Phys.\ Rev.\ Lett.\  {\bf 84}, 2322 (2000);
J.~L.~Feng, K.~T.~Matchev and F.~Wilczek,
Phys.\ Lett.\ B {\bf 482}, 388 (2000);
J.~L.~Feng and F.~Wilczek,
  Phys.\ Lett.\  B {\bf 631}, 170 (2005);
U.~Chattopadhyay, T.~Ibrahim and D.~P.~Roy,
  Phys.\ Rev.\  D {\bf 64}, 013004 (2001);
U.~Chattopadhyay, A.~Datta, A.~Datta, A.~Datta and D.~P.~Roy,
  Phys.\ Lett.\  B {\bf 493}, 127 (2000);
 S.~P.~Das, A.~Datta, M.~Guchait, M.~Maity and S.~Mukherjee,
  Eur.\ Phys.\ J.\  C {\bf 54}, 645 (2008), [arXiv:0708.2048 [hep-ph]].

\bibitem{dd_msugra}                                
  H.~Baer, V.~Barger and A.~Mustafayev,
  Phys.\ Rev.\ D {\bf 85}, 075010 (2012)
  [arXiv:1112.3017 [hep-ph]];
  J.~Ellis and K.~A.~Olive,
  Eur.\ Phys.\ J.\ C {\bf 72}, 2005 (2012)
  [arXiv:1202.3262 [hep-ph]];
  O.~Buchmueller, R.~Cavanaugh, M.~Citron, A.~De Roeck, M.~J.~Dolan, J.~R.~Ellis, H.~Flacher and S.~Heinemeyer {\it et al.},
  Eur.\ Phys.\ J.\ C {\bf 72}, 2243 (2012)
  [arXiv:1207.7315];
  O.~Buchmueller, R.~Cavanaugh, A.~De Roeck, M.~J.~Dolan, J.~R.~Ellis, H.~Flacher, S.~Heinemeyer and G.~Isidori {\it et al.},
  Eur.\ Phys.\ J.\ C {\bf 74}, 2922 (2014)
  [arXiv:1312.5250 [hep-ph]].

\bibitem{dmsugra_recent}
M. Citron {\it et al.}, \PRD(87,036012,2013), [\href{http://arXiv.org/abs/1212.2886}{arXiv:1212.2886}];
K. Kowalska, L. Roszkowski, E. M. Sessolo, \JHEP(06,078,2013), [\href{http://arXiv.org/abs/1302.5956}{arXiv:1302.5956}];
S. H.-Versille {\it et al.}, [\href{http://arXiv.org/abs/1309.6958}{arXiv:1309.6958}];
P. Bechtle {\it et al.},  [\href{http://arXiv.org/abs/arXiv:1310.3045}{arXiv:1310.3045}];
J.~Ellis, [\href{http://arXiv.org/abs/1312.5426}{arXiv:1312.5426}];
 L.~Roszkowski, E.~M.~Sessolo and A.~J.~Williams,
  JHEP {\bf 1408}, 067 (2014)
  [arXiv:1405.4289 [hep-ph]].                                                            

\bibitem{etcEllis:1985jn}
  J.~R.~Ellis, K.~Enqvist, D.~V.~Nanopoulos and K.~Tamvakis,
  Phys.\ Lett.\  B {\bf 155}, 381 (1985);
M. Drees, {\it ibid.} {\bf 158B}, 409 (1985).

\bibitem{NathMixedRep}
  A.~Corsetti and P.~Nath,
  Phys.\ Rev.\  D {\bf 64}, 125010 (2001).

\bibitem{Chattopadhyay:2001mj}
  U.~Chattopadhyay and P.~Nath,
  Phys.\ Rev.\  D {\bf 65}, 075009 (2002)
  [arXiv:hep-ph/0110341].
\bibitem{nugminter}
G. Anderson, C.H. Chen, J.F. Gunion, J. Lykken, T. Moroi, and
Y. Yamada, hep-ph/9609457; G. Anderson, H. Baer, C.H.
Chen, P. Quintana and X. Tata, Phys. Rev. {\bf D61}, 095005 (2000);
K. Huitu, Y. Kawamura, T. Kobayashi, and K. Puolamaki,
Phys. Rev. {\bf D61}, 035001 (1999);  
J.~Chakrabortty and A.~Raychaudhuri,
  Phys.\ Lett.\ B {\bf 673}, 57 (2009)
  [arXiv:0812.2783 [hep-ph]]; 
S.~P.~Martin,
  Phys.\ Rev.\ D {\bf 79}, 095019 (2009)
  [arXiv:0903.3568 [hep-ph]].


\bibitem{ucdphiggsino}
 U.~Chattopadhyay and D.~P.~Roy,
  Phys.\ Rev.\  D {\bf 68}, 033010 (2003).

\bibitem{nugmvarious}
U.~Chattopadhyay, A.~Corsetti and P.~Nath,
  Phys.\ Rev.\  D {\bf 66}, 035003 (2002);
U.~Chattopadhyay, D.~Choudhury and D.~Das,
  Phys.\ Rev.\  D {\bf 72}, 095015 (2005);               
 K.~Huitu, J.~Laamanen, P.~N.~Pandita and S.~Roy,
  Phys.\ Rev.\  D {\bf 72}, 055013 (2005);
G.~Belanger, F.~Boudjema, A.~Cottrant, A.~Pukhov and A.~Semenov,
  Nucl.\ Phys.\  B {\bf 706}, 411 (2005)
  [arXiv:hep-ph/0407218];                                  
S.~F.~King, J.~P.~Roberts and D.~P.~Roy,
  JHEP {\bf 0710}, 106 (2007)
  [arXiv:0705.4219 [hep-ph]];
S.~Bhattacharya, A.~Datta and B.~Mukhopadhyaya,
  JHEP {\bf 0710}, 080 (2007);
  K.~Huitu, R.~Kinnunen, J.~Laamanen, S.~Lehti, S.~Roy and T.~Salminen,
  Eur.\ Phys.\ J.\  C {\bf 58}, 591 (2008)
  [arXiv:0808.3094 [hep-ph]];
S.~Bhattacharya and J.~Chakrabortty,
  Phys.\ Rev.\ D {\bf 81}, 015007 (2010)
  [arXiv:0903.4196 [hep-ph]].

\bibitem{Chattopadhyay:2009fr} 
  U.~Chattopadhyay, D.~Das and D.~P.~Roy,
  Phys.\ Rev.\ D {\bf 79}, 095013 (2009)
  [arXiv:0902.4568 [hep-ph]].


\bibitem{Guchait:2011andothers}
  M.~Guchait, D.~P.~Roy and D.~Sengupta,
  Phys.\ Rev.\ D {\bf 85}, 035024 (2012)
  [arXiv:1109.6529 [hep-ph]];  
S.~Mohanty, S.~Rao and D.~P.~Roy,
  JHEP {\bf 1211}, 175 (2012)
  [arXiv:1208.0894 [hep-ph]]; 
  S.~Mohanty, S.~Rao and D.~P.~Roy,
  JHEP {\bf 1309}, 027 (2013)
  [arXiv:1303.5830 [hep-ph]];
  J.~Chakrabortty, S.~Mohanty and S.~Rao,
  JHEP {\bf 1402}, 074 (2014)
  [arXiv:1310.3620 [hep-ph]];
S.~P.~Martin,
  Phys.\ Rev.\ D {\bf 89}, no. 3, 035011 (2014)
  [arXiv:1312.0582 [hep-ph]];
S.~P.~Das, M.~Guchait and D.~P.~Roy,
Phys. Rev. {\bf D 90}, 055011 (2014),  
  arXiv:1406.6925 [hep-ph];
  I.~Gogoladze, F.~Nasir, Q.~Shafi and C.~S.~Un,
  Phys.\ Rev.\ D {\bf 90}, 035008 (2014)
  [arXiv:1403.2337 [hep-ph]].
\bibitem{Chattopadhyay:2005mv} 
  U.~Chattopadhyay, D.~Choudhury, M.~Drees, P.~Konar and D.~P.~Roy,
  Phys.\ Lett.\ B {\bf 632}, 114 (2006)
  [hep-ph/0508098].
\bibitem{nugmfinetuning}
 A.~Kaminska, G.~G.~Ross and K.~Schmidt-Hoberg,
  JHEP {\bf 1311}, 209 (2013)
  [arXiv:1308.4168 [hep-ph]];
K.~Kowalska, L.~Roszkowski, E.~M.~Sessolo and S.~Trojanowski,
  JHEP {\bf 1404}, 166 (2014)
  [arXiv:1402.1328 [hep-ph]].



\bibitem{Komine} 
  S.~Komine and M.~Yamaguchi,
  Phys.\ Rev.\ D {\bf 63}, 035005 (2001)
  [hep-ph/0007327].


\bibitem{suspect}
 A.~Djouadi, J.~-L.~Kneur and G.~Moultaka,
  Comput.\ Phys.\ Commun.\  {\bf 176}, 426 (2007)
  [hep-ph/0211331].


                                            
\bibitem{Hahn:2009zz} 
  T.~Hahn, S.~Heinemeyer, W.~Hollik, H.~Rzehak and G.~Weiglein,
  Comput.\ Phys.\ Commun.\  {\bf 180}, 1426 (2009).


\bibitem{higgsuncertainty3GeV}
 G.~Degrassi, S.~Heinemeyer, W.~Hollik, P.~Slavich and G.~Weiglein,
  Eur.\ Phys.\ J.\ C {\bf 28}, 133 (2003) [\href{http://arxiv.org/abs/hep-ph/0212020}{arxiv:hep-ph/0212020}];
 B.~C.~Allanach, A.~Djouadi, J.~L.~Kneur, W.~Porod and P.~Slavich,
  JHEP {\bf 09}, 044 (2004) [\href{http://arxiv.org/abs/hep-ph/0406166}{arxiv:hep-ph/0406166}];
 S.~P.~Martin,
  Phys.\ Rev.\ D {\bf 75}, 055005 (2007)
[\href{http://arxiv.org/abs/hep-ph/0701051}{arxiv:hep-ph/0701051}];
R.~V.~Harlander, P.~Kant, L.~Mihaila and M.~Steinhauser,
  Phys.\ Rev.\ Lett.\  {\bf 100}, 191602 (2008)
  [Phys.\ Rev.\ Lett.\  {\bf 101}, 039901 (2008)]
  [arXiv:0803.0672 [hep-ph]];
  S.~Heinemeyer, O.~Stal and G.~Weiglein,   Phys.\ Lett.\ B {\bf 710}, 201 (2012)
[\href{http://arXiv.org/abs/arXiv:1112.3026}{arXiv:1112.3026}];
A.~Arbey, M.~Battaglia, A.~Djouadi and F.~Mahmoudi,
[\href{http://arXiv.org/abs/arXiv:1207.1348}{arXiv:1207.1348}].
\bibitem{Chattopadhyay:2014gfa} 
See for example the following and references therein:
  U.~Chattopadhyay and A.~Dey,
  arXiv:1409.0611 [hep-ph] (to appear in JHEP).
\bibitem{Aaij:2013aka}
  R.~Aaij {\it et al.}  [LHCb Collaboration],
  Phys.\ Rev.\ Lett.\  {\bf 111}, 101805 (2013)
  [arXiv:1307.5024 [hep-ex]].
\bibitem{Chatrchyan:2013bka}
  S.~Chatrchyan {\it et al.}  [CMS Collaboration],
  Phys.\ Rev.\ Lett.\  {\bf 111}, 101804 (2013)
  [arXiv:1307.5025 [hep-ex]].

\bibitem{CMSandLHCbCollaborations:2013pla}
  CMS and LHCb Collaborations [CMS and LHCb Collaboration],
  CMS-PAS-BPH-13-007.

\bibitem{bsmumuextra}
  D.~Feldman, Z.~Liu and P.~Nath,
  Phys.\ Rev.\ D {\bf 81}, 117701 (2010)
  [arXiv:1003.0437 [hep-ph]];
  S.~Akula, D.~Feldman, P.~Nath and G.~Peim,
  Phys.\ Rev.\ D {\bf 84}, 115011 (2011)
  [arXiv:1107.3535 [hep-ph]].

\bibitem{btosgammalimits}
  Y.~Amhis {\it et al.}  [Heavy Flavor Averaging Group Collaboration],
  arXiv:1207.1158 [hep-ex].
\bibitem{bsgammaextra}
 For discussions on ${\rm Br}(B \rightarrow X_s \gamma)$~ a partial list is as follows:
  B.~Bhattacherjee, M.~Chakraborti, A.~Chakraborty, U.~Chattopadhyay, D.~Das and D.~K.~Ghosh,
  Phys.\ Rev.\ D {\bf 88}, no. 3, 035011 (2013)
  [arXiv:1305.4020 [hep-ph]];
  U.~Haisch and F.~Mahmoudi,
  JHEP {\bf 1301}, 061 (2013)
  [arXiv:1210.7806 [hep-ph]];
N.~Chen, D.~Feldman, Z.~Liu and P.~Nath,
  Phys.\ Lett.\ B {\bf 685}, 174 (2010)
  [arXiv:0911.0217 [hep-ph]];
M.~E.~Gomez, T.~Ibrahim, P.~Nath and S.~Skadhauge,
  Phys.\ Rev.\ D {\bf 74}, 015015 (2006)
  [hep-ph/0601163];
 U.~Chattopadhyay and P.~Nath,
  Phys.\ Rev.\ D {\bf 65}, 075009 (2002)
  [hep-ph/0110341].
\bibitem{wmap}
  G.~Hinshaw {\it et al.}  [WMAP Collaboration],
  Astrophys.\ J.\ Suppl.\  {\bf 208}, 19 (2013)
  [arXiv:1212.5226 [astro-ph.CO]].
\bibitem{planck}
  P.~A.~R.~Ade {\it et al.}  [Planck Collaboration],
  Astron.\ Astrophys.\  (2014)
  [arXiv:1303.5076 [astro-ph.CO]].


\bibitem{clic}
  A.~Sailer,
  EPJ Web Conf.\  {\bf 70}, 00085 (2014).

\bibitem{Arnowitt:1992qp} 
  R.~L.~Arnowitt and P.~Nath,
  Phys.\ Rev.\ D {\bf 46}, 3981 (1992).



\bibitem{prospino}
  W.~Beenakker, R.~Hopker, M.~Spira and P.~M.~Zerwas,
  Nucl.\ Phys.\ B {\bf 492}, 51 (1997)
  [hep-ph/9610490]; 
  W.~Beenakker, R.~Hopker and M.~Spira,
  hep-ph/9611232.

\bibitem{Baer:2012vr}
  H.~Baer, V.~Barger, A.~Lessa and X.~Tata,
  Phys.\ Rev.\ D {\bf 86}, 117701 (2012)
  [arXiv:1207.4846 [hep-ph]].
\bibitem{Djouadi:2006bz} 
  A.~Djouadi, M.~M.~Muhlleitner and M.~Spira,
  Acta Phys.\ Polon.\ B {\bf 38}, 635 (2007)
  [hep-ph/0609292].
\bibitem{SM3and4tops} 
 V.~Barger, W.~Y.~Keung and B.~Yencho,
  Phys.\ Lett.\ B {\bf 687}, 70 (2010)
  [arXiv:1001.0221 [hep-ph]]; 
  V.~Khachatryan {\it et al.}  [CMS Collaboration],
  JHEP {\bf 1411}, 154 (2014)
  [arXiv:1409.7339 [hep-ex]];
 J.~Keaveney,
  arXiv:1412.4641 [hep-ex].

\bibitem{Hisano:2004pv} 
  J.~Hisano, S.~Matsumoto, M.~M.~Nojiri and O.~Saito,
  Phys.\ Rev.\ D {\bf 71}, 015007 (2005)
  [hep-ph/0407168].
\bibitem{Belanger:2013oya} 
  G.~Belanger, F.~Boudjema, A.~Pukhov and A.~Semenov,
Comput.\ Phys.\ Commun.\  {\bf 185}, 960 (2014)
[arXiv:1305.0237 [hep-ph]].

\bibitem{xenon100}XENON100 Collaboration, E. Aprile {\it et al.}, Phys. Rev. Lett. 109, 181301 (2012).
\bibitem{lux}   
  D.~S.~Akerib {\it et al.}  [LUX Collaboration],
nd Research Facility,''                                                         
  Phys.\ Rev.\ Lett.\  {\bf 112}, 091303 (2014)
  [arXiv:1310.8214 [astro-ph.CO]].

\bibitem{xenon1t}  E.~Aprile [XENON1T Collaboration],
  Springer Proc.\ Phys.\  {\bf 148}, 93 (2013)
  [arXiv:1206.6288 [astro-ph.IM]].

\bibitem{Barger:2008qd} 
  V.~Barger, W.~Y.~Keung and G.~Shaughnessy,
  Phys.\ Rev.\ D {\bf 78}, 056007 (2008)
  [arXiv:0806.1962 [hep-ph]].
\bibitem{Icecube:2012} 
  M.~G.~Aartsen {\it et al.}  [IceCube Collaboration],
  Phys.\ Rev.\ Lett.\  {\bf 110}, no. 13, 131302 (2013)
  [arXiv:1212.4097 [astro-ph.HE]].
\bibitem{IceCube:2011aj} 
  R.~Abbasi {\it et al.}  [IceCube Collaboration],
  Phys.\ Rev.\ D {\bf 85}, 042002 (2012)
  [arXiv:1112.1840 [astro-ph.HE]].
\bibitem{Cushman:2013} 
  P.~Cushman, C.~Galbiati, D.~N.~McKinsey, H.~Robertson, T.~M.~P.~Tait, D.~Bauer, A.~Borgland and B.~Cabrera {\it et al.},
  arXiv:1310.8327 [hep-ex].
\bibitem{coupp} 
  E.~Behnke {\it et al.}  [COUPP Collaboration],
  Phys.\ Rev.\ D {\bf 86}, 052001 (2012)
  [arXiv:1204.3094 [astro-ph.CO]].
\bibitem{fermi-lat-gamma}
M.~Ackermann {\it et al.}  [Fermi-LAT Collaboration],
  Phys.\ Rev.\ Lett.\  {\bf 107} (2011) 241302
\bibitem{Ibarra:2014} 
  A.~Ibarra, M.~Totzauer and S.~Wild,
  JCAP {\bf 1404}, 012 (2014)
  [arXiv:1402.4375 [hep-ph]].

\bibitem{Gould} 
  A.~Gould,
  Astrophys.\ J.\  {\bf 321}, 560 (1987);
  A.~Gould,
  Astrophys.\ J.\  {\bf 321}, 571 (1987).

\bibitem{Griest:1986} 
  K.~Griest and D.~Seckel,
  Nucl.\ Phys.\ B {\bf 283}, 681 (1987)
  [Erratum-ibid.\ B {\bf 296}, 1034 (1988)].

\bibitem{Buckley:2013} 
  J.~Buckley, D.~F.~Cowen, S.~Profumo, A.~Archer, M.~Cahill-Rowley, R.~Cotta, S.~Digel and A.~Drlica-Wagner {\it et al.},
  arXiv:1310.7040 [astro-ph.HE].

\bibitem{Wikstrom:2009} 
  G.~Wikstrom and J.~Edsjo,
  JCAP {\bf 0904}, 009 (2009)
  [arXiv:0903.2986 [astro-ph.CO]].

\bibitem{Chacko} 
  P.~Agrawal, Z.~Chacko, C.~Kilic and R.~K.~Mishra,
  arXiv:1003.5905 [hep-ph].


\end{thebibliography}
\end{document}